\documentclass[twocolumn]{aastex701}

\usepackage{amsmath}
\usepackage{amssymb}
\usepackage{mathtools}
\usepackage{longtable}
\usepackage{tablefootnote}
\usepackage{array}

\defcitealias{Kessler2025}{K25}

\def\code#1{\texttt{#1}}

\begin{document}

\title{Chromatic Effects across the Roman Focal Plane: Implications for Supernova Photometry and Measurements of Cosmological Parameters}

\correspondingauthor{Rujuta Purohit}

\email{rujuta01@bu.edu}

\author[0009-0003-6728-9291]{Rujuta~Purohit}
\email{rujuta01@bu.edu}
\affil{Institute for Astrophysical Research \& Department of Astronomy, Boston University, 725 Commonwealth Avenue, Boston, MA 02215, USA}

\author[0000-0001-5201-8374]{Dillon~Brout} 
\email{dbrout@bu.edu}
\affil{Institute for Astrophysical Research \& Department of Astronomy, Boston University, 725 Commonwealth Avenue, Boston, MA 02215, USA}
\affil{Department of Physics, Boston University, 725 Commonwealth Avenue, Boston, MA 02215, USA}

\author[0000-0002-4934-5849]{Daniel~Scolnic}
\affiliation{Department of Physics, Duke University, Durham, NC 27708, USA}
\email{none@none.edu}

\author[0000-0003-3221-0419]{Richard~Kessler}
\affiliation{Kavli Institute for Cosmological Physics, University of Chicago, Chicago, IL 60637, USA}
\email{none@none.edu}

\author[0000-0003-0408-366X]{Jillian~Paulin}
\email{none@none.edu}
\affiliation{Department of Physics and Astronomy, University of Pennsylvania, 209 South 33rd Street, Philadelphia, PA 19104, USA}

\author[]{Stefano~Casertano}
\affiliation{Space Telescope Science Institute, 3700 San Martin Drive, Baltimore, MD 21218, USA}
\email{none@none.edu}

\author[0000-0001-5399-0114]{Nora~F.~Sherman} 
\email{none@none.edu}
\affil{Institute for Astrophysical Research, Boston University, 725 Commonwealth Avenue, Boston, MA 02215, USA}

\author[0000-0002-0476-4206]{Rebekah~Hounsell}
\affiliation{University of Maryland Baltimore County, 1000 Hilltop Cir, Baltimore, MD 21250, USA}
\affiliation{NASA GSFC, 8800 Greenbelt Rd, Greenbelt, MD 20771, USA}
\email{none@none.edu}

\author[0000-0002-2414-6886]{Eric~R.~Switzer}
\affiliation{NASA GSFC, 8800 Greenbelt Rd, Greenbelt, MD 20771, USA}
\email{none@none.edu}

\author[0000-0002-1873-8973]{Benjamin~M.~Rose}
\email{none@none.edu}
\affiliation{Department of Physics and Astronomy, Baylor University, One Bear Place \#97316, Waco, TX 76798-7316, USA}

\author[0000-0003-2764-7093]{Masao~Sako}
\email{none@none.edu}
\affiliation{Department of Physics and Astronomy, University of Pennsylvania, 209 South 33rd Street, Philadelphia, PA 19104, USA}
\affiliation{Kavli IPMU (WPI), UTIAS, The University of Tokyo, Kashiwa, Chiba 277-8583, Japan}

\author[0000-0003-0183-451X]{Lauren~Aldoroty} 
\email{none@none.edu}
\affil{University of Maryland Baltimore County, 1000 Hilltop Cir, Baltimore, MD 21250, USA}
\affil{NASA GSFC, 8800 Greenbelt Rd, Greenbelt, MD 20771, USA}

\author[0000-0001-8791-7273]{Kene~Anumba}
\email{none@none.edu}
\affiliation{Department of Physics, Duke University, Durham, NC 27708, USA}

\author[0000-0002-5389-7961]{Maria~Acevedo} 
\email{none@none.edu}
\affiliation{Department of Physics, Duke University, Durham, NC 27708, USA}

\author[0000-0001-5402-4647]{David~Rubin}
\email{none@none.edu}
\affiliation{Department of Physics and Astronomy, University of Hawai’i at Manoa, Honolulu, Hawai’i 96822, USA}
\affiliation{Lawrence Berkeley National Laboratory, 1 Cyclotron Rd., Berkeley, CA 94720, USA}

\author[0000-0003-0894-1588]{Russell~E.~Ryan~Jr.}
\email{none@none.edu}
\affiliation{Space Telescope Science Institute, 3700 San Martin Drive, Baltimore, MD 21218, USA}

\author[]{The Roman Supernova Cosmology Project Infrastructure Team}
\affil{}
\email{none@none.edu}

\begin{abstract}
    Calibration uncertainties are the leading systematics in cosmological analyses using Type Ia supernovae (SNe Ia). For the Nancy Grace Roman Space Telescope (Roman), we quantify the impact of chromatic effects on SNe Ia photometry and derived cosmological parameters, using simulated light curves from the High-Latitude Time Domain Survey. We investigate two sources of wavelength-dependent bias: focal plane array (FPA)-dependent wavelength shifts arising from spatial variations across Roman's 18 detectors, and coherent wavelength shifts corresponding to the measured $0.06\%$ uncertainty in absolute filter wavelength calibration. We probe the impact of chromatic effects by employing detector-specific filter curves that recover unbiased cosmological constraints. Using simulated SNe Ia light curves, we find that the FPA-dependent shifts---which range from +6 to -80 $\rm \AA$---introduce a redshift-dependent bias in distance modulus that, if left uncorrected, propagates to $\Delta w_0 \sim -0.06$ and $\Delta w_a = 0.236$, which are larger than the forecast statistical uncertainties of $\sigma_{\rm stat, w_0} = 0.025$ and $\sigma_{\rm stat, w_a} = 0.114$, rendering the survey systematics-limited. In contrast, a coherent 0.06\% offset in filter wavelength calibration---ranging from -3 to -11 $\rm \AA$---produces negligible redshift-dependent bias, with a minimal spread in $w_a$ ($\Delta w_a = -0.0004, \sigma_{w_{a,\rm stat}} = 0.114)$, demonstrating that the achieved prelaunch calibration precision is sufficient for this systematic to remain subdominant. Our results establish that chromatic effects are a required component of SN Ia cosmology with Roman.
\end{abstract}

\keywords{\uat{Cosmology}{343} --- \uat{Type Ia supernovae}{1728}}

\section{Introduction}
\label{sec:intro}

The discovery of the accelerating expansion of the universe \citep[][]{Riess1998, Perlmutter1999} has motivated an era of large-scale surveys with the goal of measuring cosmological parameters with unprecedented precision. Designed as a Stage IV cosmology experiment, the Nancy Grace Roman Space Telescope (Roman) is NASA’s next planned flagship mission \citep[][]{Spergel2013, Spergel2015, Akeson2019},launched on August 30, 2026. Roman features a wide-field instrument (WFI) which has 8 imaging filters, a prism, and a grism. The WFI has 18 Teledyne H4RG-10 sensor chip assemblies \citep[SCAs;][]{Mosby2020} covering a 0.281 deg$^2$ field of view and broadband filters with central wavelengths 6200 (F062/R), 8690 (F087/Z), 10600 (F106/Y), 12930 (F129/J), 14640 (F146/W), 15770 (F158/H), 18420 (F184/F), and 21250 (F213/K) $\rm \AA$ respectively. Roman will carry out its cosmology observations partially through the High-Latitude Time Domain Survey (HLTDS), which is a core community survey designed to discover more than 20,000 Type Ia supernovae (SNe Ia) out to $z \lesssim 3$, for which $\sim 10,000$ satisfy selection requirements for a cosmology analysis \citep[][]{Hounsell2018, Rose2021, Kessler2025, Rubin2025a}. In this work we adopt the most up-to-date HLTDS observing strategy as implemented in the new ``Sundial" cadence and simulation strategy \citep[][]{Paulin2026}.

One of Roman's primary scientific goals is to probe the nature of dark energy with SNe Ia. SNe Ia are ``standardizable candles.'' The light curves of SNe Ia can be used to estimate the distance to each event, and comparing these distances across redshifts yields estimates of physical quantities such as the expansion rate $H_0$, dark energy $\Omega_{\Lambda}$, and equation-of-state parameter of dark energy \textit{w}. A number of ground-based SN surveys and reanalyses of public datasets have significantly improved the statistical and systematic precision in recent years. Several recent analyses report a $2\sigma-3\sigma$ deviation from $\Lambda$CDM and point towards an evolving dark energy model \citep[e.g., Dark Energy Survey (DES) and Union3;][]{Brout2022, DES2024, Rubin2025a, Popovic2025}, consistent with independent evidence from measurements of baryonic acoustic oscillations by the Dark Energy Spectroscopic Instrument \citep[DESI;][]{Adame2025}.

\begin{figure}[t!]
    \centering
    \includegraphics[width=\columnwidth]{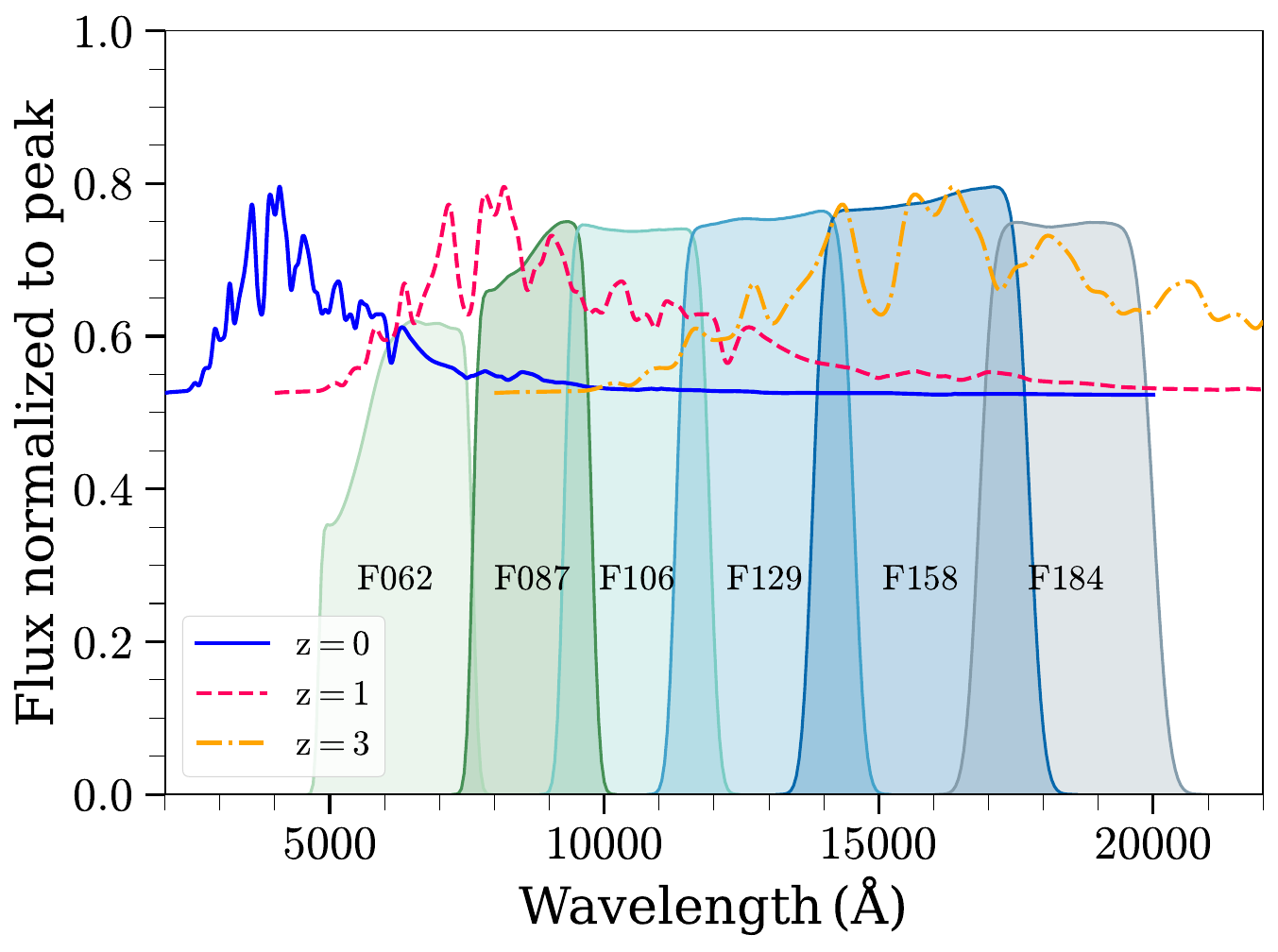}
    \caption{Standard SN Ia spectrum at redshifts $z = 0, 1,$ and $3$ with the Roman broadband filters overlaid.}
    \label{fig:sn_spectrum}
\end{figure}

When conducting cosmological analyses with SNe Ia, calibration of the photometric system is typically one of the dominant sources of systematic uncertainty \citep[][]{Schlafly2012, Betoule2014,Scolnic2015, Scolnic2018, Lasker2019, Jones2019, Vincenzi2024}. In fact, uncertainties in absolute calibration at the 1\% level typically result in uncertainties in the equation of state \textit{w} for dark energy of the order of 0.03 \citep[][]{Brout2022}, and calibration systematics remain a key consideration in recent analyses reporting evidence for evolving dark energy. Because calibration errors are survey- and filter-dependent, and propagate through both SN Ia model training \citep[e.g.,][]{Taylor2021} and light-curve fitting, they naturally manifest as redshift-dependent biases in inferred distances. These uncertainties primarily affect cosmological constraints through two closely related mechanisms. First, cosmological constraints depend on comparing the relative brightnesses of SNe Ia at different redshifts. As the rest-frame SN Ia spectrum is redshifted, we observe it in different bandpasses, which must be calibrated relative to each other (see Figure \ref{fig:sn_spectrum}). Any errors in this cross-band photometry propagate directly into distance estimates as a function of redshift. Second, SNe Ia are observed under spatially and temporally varying conditions---across different regions of the sky and positions on the focal plane.

Both of these effects (relative bandpasses at different redshifts and observing conditions) are governed by the wavelength-dependent instrumental response. Previous cosmology analyses characterize calibration systematics in terms of uncertainties in each filter's zero-point and transmission efficiency as a function of wavelength, $T(\lambda)$. Here we focus on the latter, characterized by shifts in the transmission function $T(\lambda) \rightarrow T(\lambda + \delta\lambda)$. In practice, mismatches between the true spectral energy distribution (SED) of a source and the flux measured with the instrument lead to photometric errors that depend on the object's color and redshift. We refer to these wavelength-dependent transmission shifts collectively as ``chromatic effects," and they provide a framework for understanding how photometric calibration uncertainties translate into redshift-dependent biases in SN Ia distance measurements.

Improvement in the understanding of systematic uncertainties is crucial to taking advantage of the order-of-magnitude increases in statistics expected in the coming years. For ground-based surveys, chromatic effects arise from both variations in atmospheric transmission and the instrumental spectrophotometric response. For Roman, which observes from space, atmospheric contributions are absent and the primary source of chromatic effects is the spectrophotometric response of the detectors. Discrepancies between the actual and modeled instrumental response can cause chromatic shifts in photometry, which are particularly significant when measuring objects across a range of redshifts. For Roman, weak lensing and SN Ia science require absolute photometric calibration within $2\%$ ($1\%$ goal) and relative calibration between filters to $0.5\%$, translating to a required precision in filter edge wavelength characterization of $0.06\%$ over the field of view \citep[][]{Switzer2025}. Throughout this paper, ``edge wavelengths" refer to the wavelengths where filter transmission falls below $50\%$ of peak transmission.

Chromatic effects have been observed in and dealt with for various other SN~Ia cosmology surveys. Works such as those by \citet{Lasker2019} and \citet{Lee2023} applied SED-dependent chromatic corrections to DES SN~Ia light-curve photometry, finding that the combined atmospheric and instrumental transmission function introduces biases at the 10 mmag level if left uncorrected---comparable to the dominant systematic uncertainties in calibration---with large redshift-dependent biases of $\sim$0.02 mag in distance modulus found when considering individual CCDs. More recently, \citet{Marlin2025} identified transmission-function color dependences as the largest systematic uncertainty in the ATLAS/TITAN dataset, finding that uncorrected transmission-function color dependences introduced photometric offsets at the $\sim$30 mmag level if left uncorrected, and achieved improvements in chip-to-chip calibration offsets from $\sim$17 to $\sim$3 mmag through cross calibration with DES Y6 tertiary stars. Both studies underscore that the impact of chromatic effects at the mmag level is a common challenge across modern SN~Ia surveys, and motivate the careful characterization we have performed here for Roman.

In this work, we quantify the impact of chromatic effects on Roman SN Ia cosmology by simulating realistic spatial variations in filter effective wavelengths across the WFI focal plane (hereafter ``focal plane-dependent shifts"). Using laboratory measurements of the wavelength-dependent transmission profiles for each of the 18 SCAs presented by \citet{Switzer2025}, we generate light curves for SNe Ia spanning redshifts $0.1 < z < 3$, following the HLTDS observing strategy. We perform light-curve fitting and analyze how chromatic effects affect both the single-epoch photometry and the derived cosmological parameters. Our analysis reveals systematic redshift-dependent trends in the inferred magnitudes and distance moduli, with implications for constraints on the dark energy equation-of-state parameters $w_0$ and $w_a$. By characterizing these effects in a controlled simulation framework, we provide quantitative guidance for incorporating the impact of chromatic effects into the HLTDS analysis pipeline and assess their importance relative to the mission's target systematic error budget.

This paper is organized as follows. In \S \ref{sec:methods}, we describe our simulations and the exact nature of the chromatic effects we model. In \S \ref{sec:results}, we present our results on single-epoch SN Ia photometry, nuisance parameters, and cosmology fitting. In \S \ref{sec:discussion}, we present a brief discussion of our results and explore the potential scope for future work.

\section{Methods}
\label{sec:methods}

In this section, we describe the exact form of the chromatic effects and the manner in which they are applied to the SN Ia photometry.

\subsection{Brief Overview of the Roman Space Telescope}

The WFI in Roman offers visible to near-infrared imaging and spectroscopy with a large field of view of 0.281 deg$^2$, $0.1 \arcsec$ spatial resolution, and high sensitivity for detecting faint objects. Incoming light is filtered or dispersed by the Element Wheel Assembly (EWA) before reaching the detectors. The EWA is a filter wheel with positions for 11 optical elements, including eight filters (F062, F087, F106, F129, F158, F184, F213, F146), a grism, a prism, and a dark element. As explained in \S \ref{subsec:sim} below, following the in-guide recommendation of the HLTDS committee, we omit F146 and F213 bands from our analysis and thus they are not listed in Table \ref{tab:wavelengths}\footnote{https://asd.gsfc.nasa.gov/roman/comm\_forum/forum\_17/Core \\ \_Community\_Survey\_Reports-rev03-compressed.pdf}.

\begin{figure*}[t!]
    \centering
    \includegraphics[width=\columnwidth]{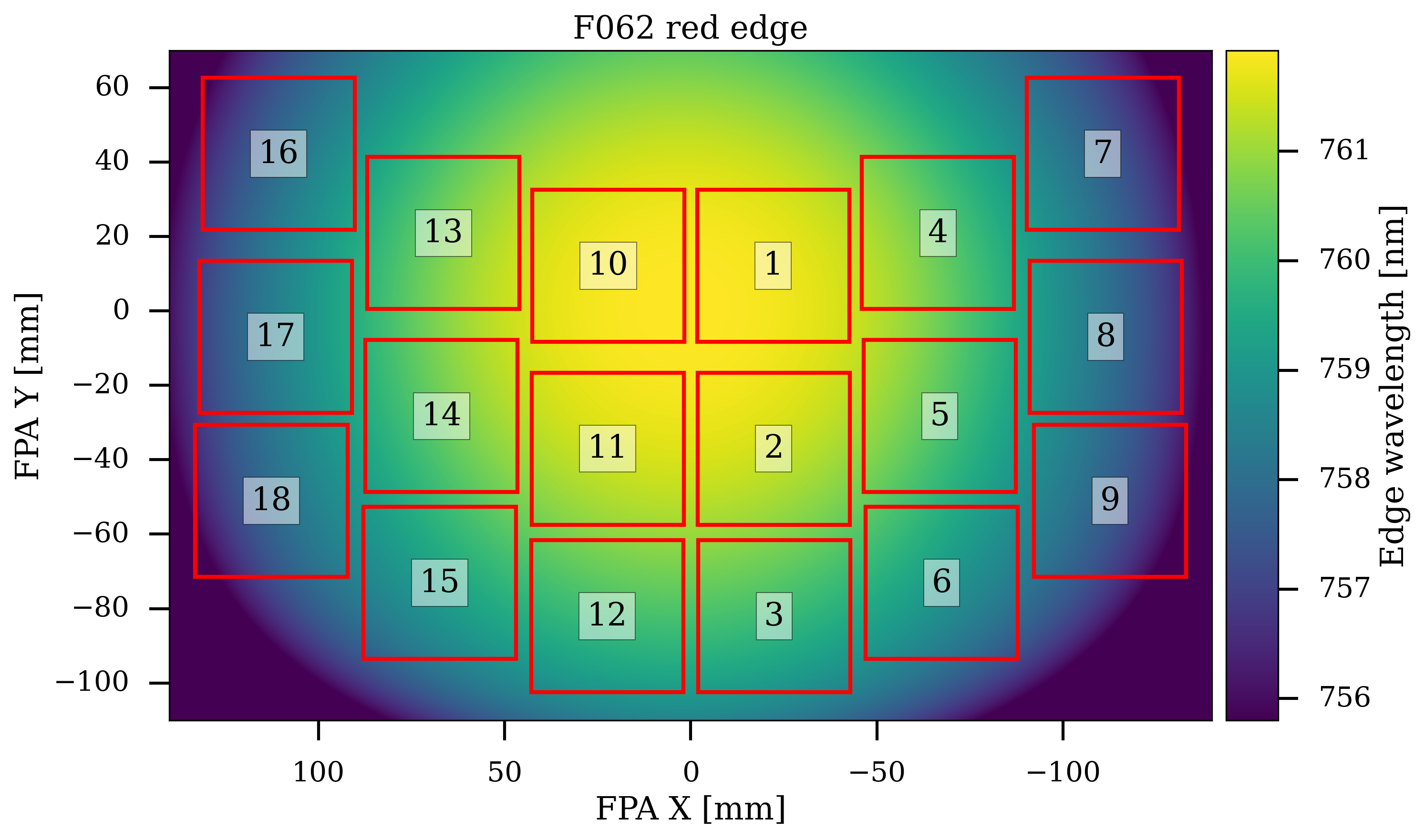}
    \includegraphics[width=\columnwidth]{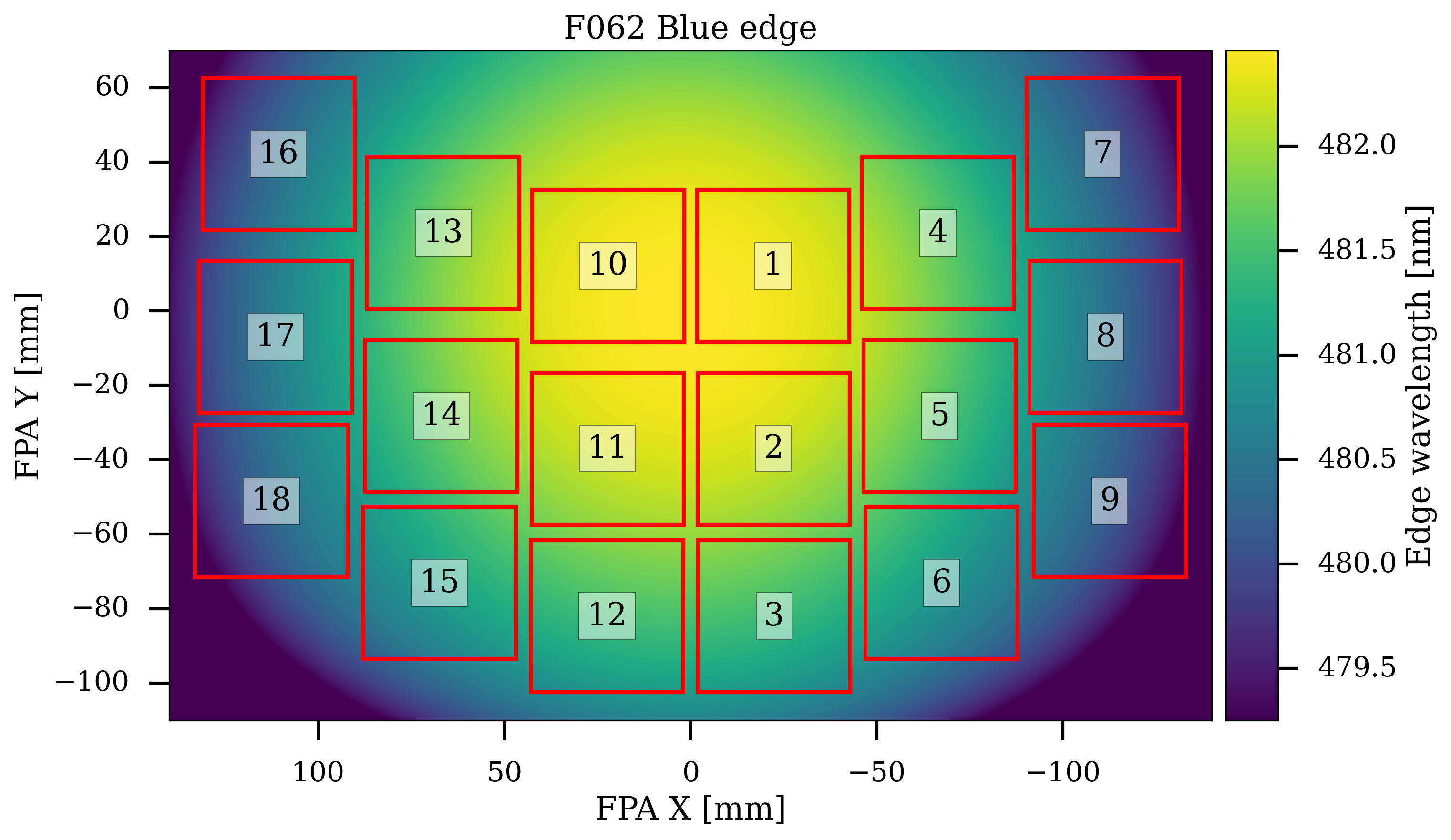}
    \caption{Filter edge wavelengths for each SCA in the F062 band. The left panel shows the red edge while the right panel shows the blue edge. These plots are created using the Roman WFI Optical Model developed in \citet{Switzer2025}.}
    \label{fig:edges}
\end{figure*}

\subsection{Focal Plane-dependent Wavelength Shifts}

The primary source of chromatic effects for Roman is the spectrophotometric response of the detectors on the WFI. The throughput for each of the Roman filters varies across the focal plane assembly (FPA) as a function of the wavelength. Due to this dependence, observing the same source at different positions on the FPA leads to different fluxes. The wavelength-dependent transmission for the Roman filters has been characterized with laboratory tests, accounting for effects of incident angle and variations in filter coating thickness \citep{Switzer2025}. Specifically, filter edges were characterized using full-field diffuser illumination, which allowed for the retrieval of the field dependence and low-order variations across the optical surface. The edge wavelengths themselves are now known with a $0.06\%$ precision. We utilize this precision as another constraint in our simulations by applying a ``coherent shift" of $0.06\%$ to the filter edge wavelengths in \S \ref{subsec:coherent}. When a filter is ``coherently" shifted, we mean that all individual SCA throughputs across the FPA are shifted by the same value for a given filter.

Ground thermal-vacuum testing shows that the field-dependent filter edge wavelengths are modeled to $0.06\%$ residuals across the field \citep{Switzer2025}, establishing the ground knowledge level adopted here. Because the ground illumination differs from the flight pupil, commissioning will provide the first end-to-end measurement of the in-flight bandpass, while subsequent flight calibration will refine the low-order model for the field dependence. Through pupil imaging, \citet{Switzer2025} directly mapped variations in filter transmission across the FPA. These maps, together with other WFI thermal-vacuum testing data, were incorporated into the semiempirical Roman WFI Optical Model\footnote{github.com/RomanSpaceTelescope/RST\_WFI\_optical\_model}, which we use to characterize the variations in throughput.

To model the wavelength-dependent transmission for each SCA, we shift the filter bandpass using measured TVAC edge wavelength data from \citet{Switzer2025}. Specifically, for each filter and SCA, we apply a linear transformation that maps the red and blue edges of the SCA 2 bandpass (taken as the nominal reference, corresponding to the center of the focal plane) to their measured TVAC positions for SCA $j$:

\begin{equation}
    \lambda_{\rm{red~edge}, \, j} = m \lambda_{\rm{red~edge}, \, 2} + C
\end{equation}

\begin{equation}
    \lambda_{\rm{blue~edge},\,  j} = m \lambda_{\rm{blue~edge}, \, 2} + C
\end{equation}

The slope $m$ and intercept $C$ are determined by simultaneously solving these two equations for each filter–SCA pair. Using the slope and intercept, we identify a linear transformation from the SCA 2 throughput ($\lambda_{\rm{nominal}}$) to an SCA $j$ throughput ($\lambda_{\rm{shift}}$), such that the red and blue edge wavelengths together match the TVAC data:

\begin{equation}
    \lambda_{\rm{shifted}} = m\lambda_{\rm{nominal}} + C
\end{equation}

\begin{figure*}[t!]
    \centering
    \includegraphics[width=2\columnwidth]{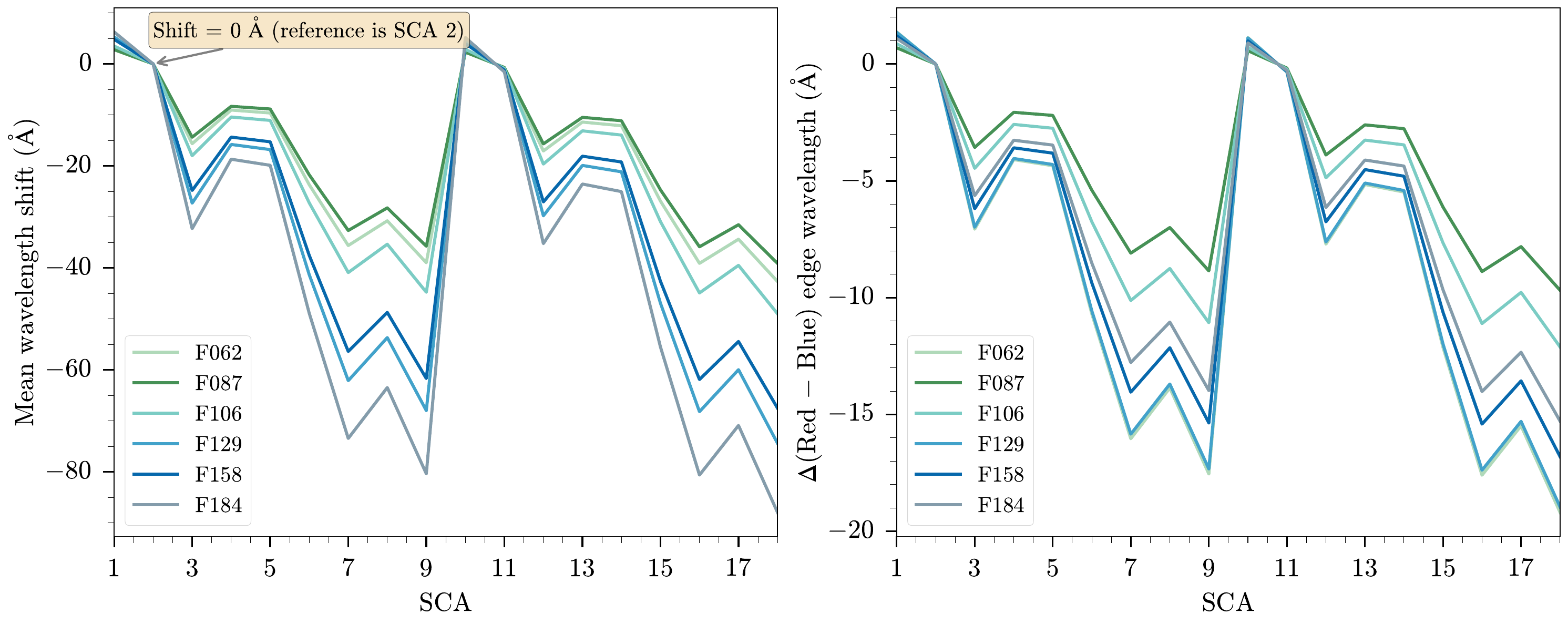}
    \caption{Left: the average wavelength shift for the FPA-dependent case in $\rm \AA$ for each SCA with respect to the center of SCA 2 (which can be seen to have a shift of 0). Right: the average difference in red and blue edges for each SCA relative to SCA 2.}
    \label{fig:shifts}
\end{figure*}

The resulting per-SCA mean wavelength shifts across all filters are shown in the left panel of Figure~\ref{fig:shifts}. 

\subsection{Coherent Shifts}
\label{subsec:coherent}

In addition to the FPA-dependent spatial variations described above, we consider the impact of a uniform systematic offset in filter wavelengths across the entire focal plane. Such ``coherent shifts" represent the residual calibration uncertainty that remains after an FPA-dependent correction has been applied — specifically, the irreducible error that arises if the spectrophotometric response model retains a systematic offset at the level of the prelaunch measurement precision. Based on the precision achieved in laboratory characterization \citep{Switzer2025}, we implement coherent shifts by applying a uniform 0.06\% offset to the central wavelength of each filter. This corresponds to absolute wavelength shifts in $\rm \AA$ of  -3.7, -5.2, -6.4, -7.8, -9.5, and -11.1 for filters F062, F087, F106, F129, F158, and F184, respectively (shown in Table.~\ref{tab:wavelengths}). For Figures \ref{fig:residuals}, \ref{fig:mag_template}, and \ref{fig:ab_coherent_fits}, we consider the scenario in which all filters shifts are systematically identically ($+0.06\%$). Hereafter, this approach is called `COH-FIXED'. We also consider a more realistic scenario in which each filter's offsets are uncorrelated (hereafter `COH-RANDOM'). To accomplish this, we draw randomly from a Gaussian distribution for each filter with a width of the shifts given above. Each set of random shifts is used in a different realization of the simulation. 

The 0.06\% uncertainty in the edge wavelength calibration from prelaunch TVAC testing is an estimate of the error on absolute filter wavelength measurement. This serves as a critical test case for Roman calibration: it is important to understand how severely cosmological constraints would be affected if the spectrophotometric response model retains a systematic offset at this level. Comparing the FPA-dependent and coherent shift scenarios thus isolates the relative importance of spatial variations versus overall systematic offsets in the chromatic effects budget. The associated average correction in magnitude required to account for the chromatic effects (``chromatic correction") for each filter is listed in Table \ref{tab:wavelengths}.

It is useful to reframe these two systematic cases in terms of a calibration hierarchy. The FPA-dependent wavelength shifts represent a known, position-dependent effect that must be modeled and corrected on a per-object basis using SCA-specific filter curves. In this sense, the FPA-dependent shifts, as implemented in this work, characterize the worst-case scenario if no FPA-dependent correction is applied and provide a framing for the level to which an uncertainty on the FPA-dependent corrections must be trusted. The coherent shifts, by contrast, represent residual errors in the spectrophotometric response model, meaning that the uncertainty that remains after an FPA-dependent correction has been applied, but the absolute wavelength scale retains a systematic offset at the premeasured 0.06\% uncertainty level. Under this interpretation, the coherent shifts characterize the impact of imperfect knowledge of the edge wavelengths, rather than a distinct physical systematic. We adopt this framing in the cosmology discussion presented in \S~\ref{sec:discussion}.

\begin{table*}
\title{Properties of the Filters Used on Roman}
\centering
\renewcommand{\arraystretch}{1.2}
\setlength{\tabcolsep}{5pt}
\begin{tabular}{cccccc}
\hline
\hline
& & & & Mean & \\
& & & & COH- & COH-\\
& & Mean & FPA & FIXED- & FIXED-\\
Bandpass & Central $\lambda$ & FPA $\Delta \lambda$ & Impact & Shift & Impact\\
 & ($\rm \AA$) & ($\rm \AA$) & (mmag) &  ($\rm \AA$) & (mmag)\\
(1) & (2) & (3) & (4) &  (5) & (6)\\
\hline
\hline
F062 & 6200  & -19.01 & -11.9 & -3.7  & -2.3 \\
F087 & 8690  & -17.44 & -7.3  & -5.2  & -2.1 \\
F106 & 10600 & -21.85 & -17.8 & -6.4  & -5.3 \\
F129 & 12930 & -33.16 & -14.0 & -7.8  &  3.3 \\
F158 & 15770 & -30.11 & -7.5  & -9.5  &  2.3 \\
F184 & 18420 & -39.21 &  0.4  & -11.1 & -0.1 \\
\hline
\end{tabular}
\caption{Column (1): Roman filter name. Column (2): central wavelength for nominal filter. Column (3): mean wavelength shift across all 18 SCAs for the FPA-dependent case in $\rm \AA$ defined relative to SCA 2. Column (4): average photometric impact in mmag corresponding to the mean FPA-dependent shifts across all simulated SNe Ia and epochs in that filter; this represents the average photometric offset that would arise if FPA-dependent shifts were ignored. Column (5): The 0.06\% uncertainty in the central wavelength from prelaunch TVAC testing (COH-FIXED), which corresponds to a shift in the wavelength for each filter. Column (6): the impact in mmag for the COH-FIXED case. The impact is determined for an SN Ia at peak using the SALT3 model.}
\label{tab:wavelengths}
\end{table*}

\subsection{Simulations}
\label{subsec:sim}

We utilize the SuperNova ANAlysis software \citep[\code{SNANA};][]{Kessler2009} jointly with the pipeline tool \code{Pippin} \citep{Hinton2020} to simulate SNe Ia observed as part of the HLTDS and perform a cosmology analysis. We briefly review two prior simulation efforts before describing the updated simulation used in this work.

\citet{Rose2025} (``Hourglass") presented the first comprehensive HLTDS simulation, using an early-concept observing strategy of four filters per tier at a five-day cadence, a WIDE tier of $19 \rm ~deg^2$, and a DEEP tier of $4.2 \rm ~deg^2$, with $\sim 20\%$ of the area also covered by prism observations. The simulation included 10 extragalactic time-domain classes and, with a detection threshold of signal-to-noise $\geq 5$ in two filters, projected approximately 21,700 SNe~Ia over the survey lifetime.

\citet{Kessler2025} (hereafter \citetalias{Kessler2025}) presented the first end-to-end cosmological analysis using the in-guide strategy recommended by the HLTDS Definition Committee, and this is the first such analysis to simultaneously employ both photometric classification and photometric redshifts. Using the two-year core component of the survey, they simulated four subsamples---Roman WIDE and DEEP tiers combined with the Legacy Survey of Space and Time (LSST) wide-fast-deep (WFD) and deep drilling field (DDF) samples---and carried out a detailed analysis including light-curve fitting, photometric SN+host redshifts, photometric classification, and various analyses for mitigating contamination and correcting distance bias. Following selection criteria, the resulting Hubble diagram contains $\sim 10000$ Roman SNe~Ia combined with $\sim 4400$ LSST events, yielding a dark energy figure of merit (FoM) well above the NASA mission requirement of 326. \citetalias{Kessler2025} also presents a preliminary evaluation of the pilot and extended visit components, demonstrating that extended visits increase the DEEP-tier SN~Ia sample by $\sim$8\%, and that a 20 day pilot visit cadence is optimal for accumulating an early high-redshift sample.

However, neither \citet{Rose2025} nor \citetalias{Kessler2025} incorporated the complete set of recommendations from the HLTDS Definition Committee in a single, self-consistent simulation. Specifically, neither included a comprehensive examination of the pilot and extended visit components, modeled the rotation of the telescope over the survey period, nor accounted for the reality that only $7/8$ of the FPA is filled with active pixels (\citetalias{Kessler2025} accounts for the active pixels with random rejection as a very rough approximation). Critically, for the present work, neither simulation resolved variations in filter transmission at the level of individual SCAs. Such resolution is essential for quantifying photometric systematics arising from both FPA-dependent and coherent filter wavelength shifts.

We therefore make use of ``Sundial" \citep{Paulin2026}, the first HLTDS cadence and observing strategy to implement the full Definition Committee strategy in a single unified framework. Five WFI filters are used per tier to span the rest-frame optical\footnote{See Table 1 in \citetalias{Kessler2025} for in-guide recommendation from HLTDS Definition Committee}: F062 (5 day cadence), F087, F106, F129, and F158 (10 day cadence) for the WIDE tier, and F087 (5 day cadence), F106, F129, F158, and F184 (10 day cadence) for the DEEP tier; F213 is omitted following committee recommendations. The survey comprises eight pilot visits in Year~1, a two-year core component, and eight extended visits bracketing the core (DEEP tier only). Simulated observations began on 2026 July~1 with a cadence of $5 \pm 1$ days between visits, resulting in 148 visits over two years.

Sundial identifies the individual SCA for each supernova observation, facilitating the SCA-level photometric analysis. Additionally, Sundial incorporates a substantial overlap region between the DEEP and WIDE tiers in the northern sky, which produces a significant population of matches at redshifts $z > 2.5$---events that are comparatively rare in \citetalias{Kessler2025}.

Using the cadence and observing strategy described in Sundial, we simulate observations of SNe Ia for Roman across the redshift range $0.1 < z < 3$. Light curves are generated with the SALT3 model, with stretch and color population parameters drawn from \citet{Popovic2023} and the DES-SN5YR analysis. The volumetric rate follows a broken power law. Redshifts are treated as perfectly known (e.g. from host-galaxy spectroscopy), with no simulated redshift uncertainty. Milky Way extinction is applied using the \citet{SFD1998} dust map, with no additional variation in reddening law applied beyond the fiducial Galactic extinction. We do not simulate core-collapse supernovae or peculiar Type Ia subtypes (e.g.\ 91bg-like, 91T-like) in this analysis; modeling of non-Ia contamination is studied in \citetalias{Kessler2025}.

To isolate the photometric effect of chromatic effects, we run three independent simulations---nominal (no shifts), FPA-dependent shifts, and coherent shifts---using identical random seeds so that the same SN~Ia population is drawn in each case. Differences in the recovered distance moduli therefore reflect only the effect of the chnages in filter transmission, with no statistical fluctuations from drawing different samples.

To implement FPA-dependent shifts, we record the location of each observed SN~Ia on the focal plane and assign it the transmission curve appropriate to its SCA. With 18 SCAs and six bands, this would nominally require 108 filter curves; however, \code{SNANA} currently supports at most 62 defined filters \footnote{Current \code{SNANA} infrastructure allows 62 defined filters due to the requirement of a unique single-character name (lowercase and uppercase alphabets and numbers 0 through 9).}. We use the empirical observation that only nine unique filter transmissions are needed across the 18 SCAs (pairs 1–10, 2–11, 3–12, 4–14, 5–13, 6–15, 7–18, 8–17, and 9–16 share identical curves), yielding 54 distinct filter curves that fit within the \code{SNANA} limit.

To implement COH-FIXED shifts, we create a simulation where the filter transmission functions are offset in wavelength by the fixed coherent shift defined in column~(5) of Table~\ref{tab:wavelengths} (Mean COH-FIXED shift). To implement COH-RANDOM shifts, we draw nine random offsets per filter from a Gaussian distribution centered on the nominal bandpass wavelength with a standard deviation of 0.06\%, and run a separate simulation for each realization (only one realization is needed for FPA-dependent and COH-FIXED). Each filter's offset is treated as uncorrelated with those of the other filters.

\subsection{Light Curves, Distances, and Cosmology Fitting}

Distances to SNe Ia are typically estimated by fitting their photometric light curves to an underlying spectral time-series model. The fitting outputs summary statistics that characterize each light curve: the peak magnitude $m_B$ is defined as $m_B = -2.5 \log_{10}(x_0)$, the duration stretch ($x_1$), and the color ($c$). Using these parameters, distance modulus can be computed with the Tripp estimator \citep{tripp1998}:

\begin{equation}
\label{eq:tripp}
    \mu = m_B + \alpha x_1 - \beta c - M_0
\end{equation}

\noindent where $m_B$ is the log of the fitted light-curve amplitude $x_0$, $M_0$ is the rest-frame magnitude for an SN Ia with $x_1 = c = 0$, $\alpha$ is the stretch-magnitude standardization parameter, $x_1$ is the stretch, \textit{c} is the color, and $\beta$ is the color-magnitude standardization parameter. The nuisance parameters $\alpha$ and $\beta$ are single numbers that correlate light-curve properties with an intrinsic luminosity and are determined such that they minimize the scatter in distances of SNe Ia \citep[][]{Marriner2011}. 

The rest-frame SED for each SN depends on the color and stretch parameters mentioned above. For each epoch, the SED undergoes cosmological dimming, is redshifted from the source frame to the observer frame, and is multiplied by the filter transmission functions. Wavelength-dependent Milky Way extinction is applied using $R_V = 3.1$ \citep{Fitzpatrick1999}.

This process is done first using the standard Roman filters and then using the shifted filters. To standardize the SN Ia brightnesses, we fit light curves using \code{SALT3-NIR} \citep[]{Guy2007, Pierel2022, Taylor2023}. For each SN, the SALT3 light-curve fit returns the parameters mentioned above: $m_B$, $x_1$, and $c$, along with the time at which the SN brightness peaks. 

We apply the following quality cuts to our sample:

\begin{enumerate}
    \item Convergent SALT3 fit.
    \item Peak signal-to-noise (SNR) ratio $\texttt{SNRMAX} > 10$ in at least one band, and $\texttt{SNRMAX} > 5$ in at least two additional bands, ensuring detections across a minimum of three photometric bands.
    \item $\texttt{SNRSUM} > 20$ where \texttt{SNRSUM} is the quadrature sum of SNR over observations with $ -15 < T_{\rm rest} < + 45$ days in the rest frame.
    \item Rest-frame phase of the first observation to satisfy $T_{\rm rest,min} < -5$ days and the last observation to satisfy $T_{\rm rest,max} > +15$ days, providing coverage through and beyond maximum light. 
    \item $|x_1| < 3.0$, $|c| < 0.3.$
\end{enumerate}

Relative to \citetalias{Kessler2025}, we adopt more relaxed selection criteria in order to study chromatic effects across the full SN~Ia population, not only the cosmology-grade sample; more stringent cuts (than the ones mentioned above) are imposed at the cosmology-fitting stage and bring the sample down to $\sim 10,000$ SNe. We note that our nominal simulation recovers an FoM of 624, compared to 578 in \citetalias{Kessler2025}. This improvement is driven primarily by the substantial population of high-$z$ SNe~Ia in the WIDE+DEEP overlap region, which is absent from \citetalias{Kessler2025}, and by the use of spectroscopic rather than photometric redshifts for the Roman sample, which reduces $\mu$ errors at moderate and high redshifts.

\begin{figure*}
    \centering
    \includegraphics[width=2\columnwidth]{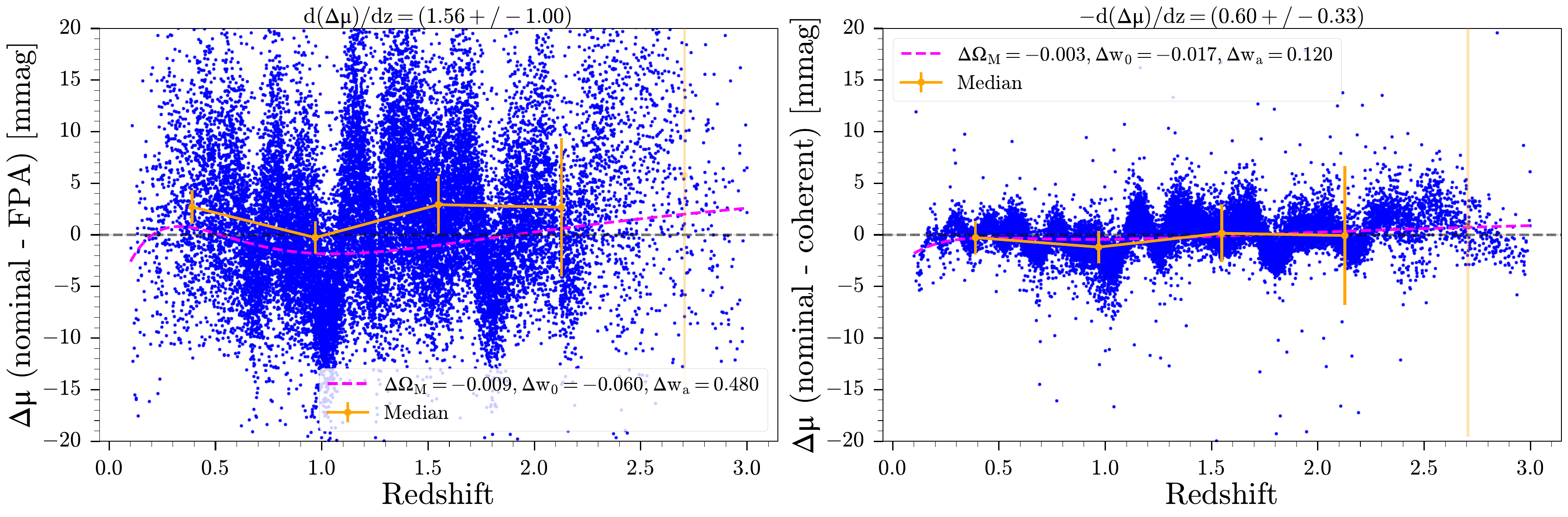}
    \caption{Hubble diagram shifts (nominal - shifted) for both our simulations. The left panel shows the FPA-dependent shifts and the right panel shows COH-FIXED shifts. The solid orange line connects the medians in each redshift bin. Since we have fewer data at $z > 2.5$, the errors are large ($\propto 1/\sqrt{N}$) and we show them in a lighter color. The dashed pink line represents the best fit $w_0 w_a$CDM cosmology calculated using a simple Roman-only sample Markov Chain Monte Carlo method. The slopes of the best-fit lines are given above each panel in millimagnitudes.}
    \label{fig:residuals}
\end{figure*}

After light-curve fitting, the next analysis stage is to use the method of BEAMS with bias corrections \citep[BBC;][]{Kessler2017, Kunz2007, Hlozek2012}, and to apply distance bias corrections for the Hubble diagram as a multidimensional function of redshift, color, and stretch \citep{Kessler2017, Popovic2021}. The nuisance parameters $\alpha$ and $\beta$ are fit separately with BBC for nominal and the two shifted simulations. The likelihood (as given in Equation 6 of \citealt{Kessler2017}) results in a cosmology-independent minimization of the free parameters, which aims to minimize the scatter in the Hubble diagram. The differences in distance modulus we obtain (between the nominal and shifted filters) are shown in Figure \ref{fig:residuals}. We also plot the best-fit cosmology we obtain at this stage (before using stringent cosmology-grade constraints, which is explained in \S \ref{sec:cosmo}).

For the FPA-dependent shifts, we find outliers (residuals $> 50$ mmag) in the plot of differences in distance modulus at $z \sim 1.0$, 1.8, and 2.4, corresponding to filter dropout transitions where F062, F087, and F106 respectively shift into the rest-frame UV and are too blue for the SALT3 model. To verify these outliers are not artifacts of the filter shifts applied, we conducted a high-SNR test by setting the exposure time equal to 100,000 for each event, which tests whether the outliers are driven by photometric noise. The outliers are not seen in this test, confirming that they result from the systematic loss of filter coverage at these transitions rather than chromatic systematics. Thus, the outliers in our simulations are statistically expected.

\begin{figure}[t!]
    \centering
    \includegraphics[width=\columnwidth]{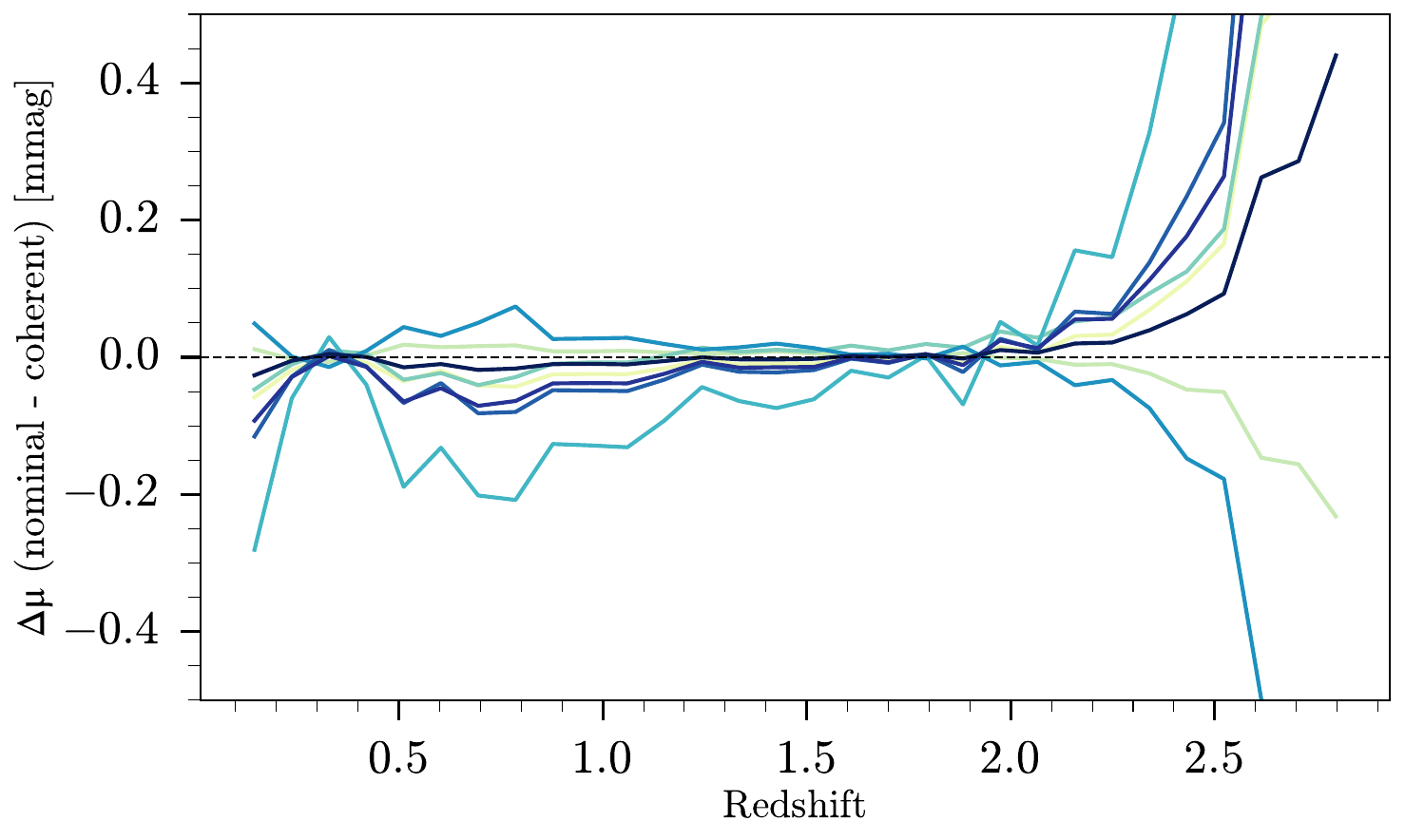}
    \caption{Shifts in Hubble diagram in millimagnitudes of the nine realizations of our COH-RANDOM simulation. We observe a smaller spread than in the the FPA-shifted case, which is consistent with the overall size of the filter shifts implemented.}
    \label{fig:9_HD}
\end{figure}

For coherent shifts we incorporate all nine realizations from COH-RANDOM to probe the overall scatter in measured cosmological parameters. The nine differences in Hubble diagram are shown in Figure \ref{fig:9_HD}. The scatter is extremely small and is constrained to $-5$ to $+5$ mmag.

\section{Results}
\label{sec:results}

In this section, we discuss the impact of uncorrected chromatic effects, and anticipated correction uncertainties on the single-epoch photometry and show their dependence on SN color, redshift, and phase. We also present the changes in light-curve fit parameters, distance moduli, and cosmological parameters due to these effects.

\subsection{Single-epoch Supernova Photometry}

\begin{figure*}[t!]
    \centering
    \includegraphics[width=1.75\columnwidth]{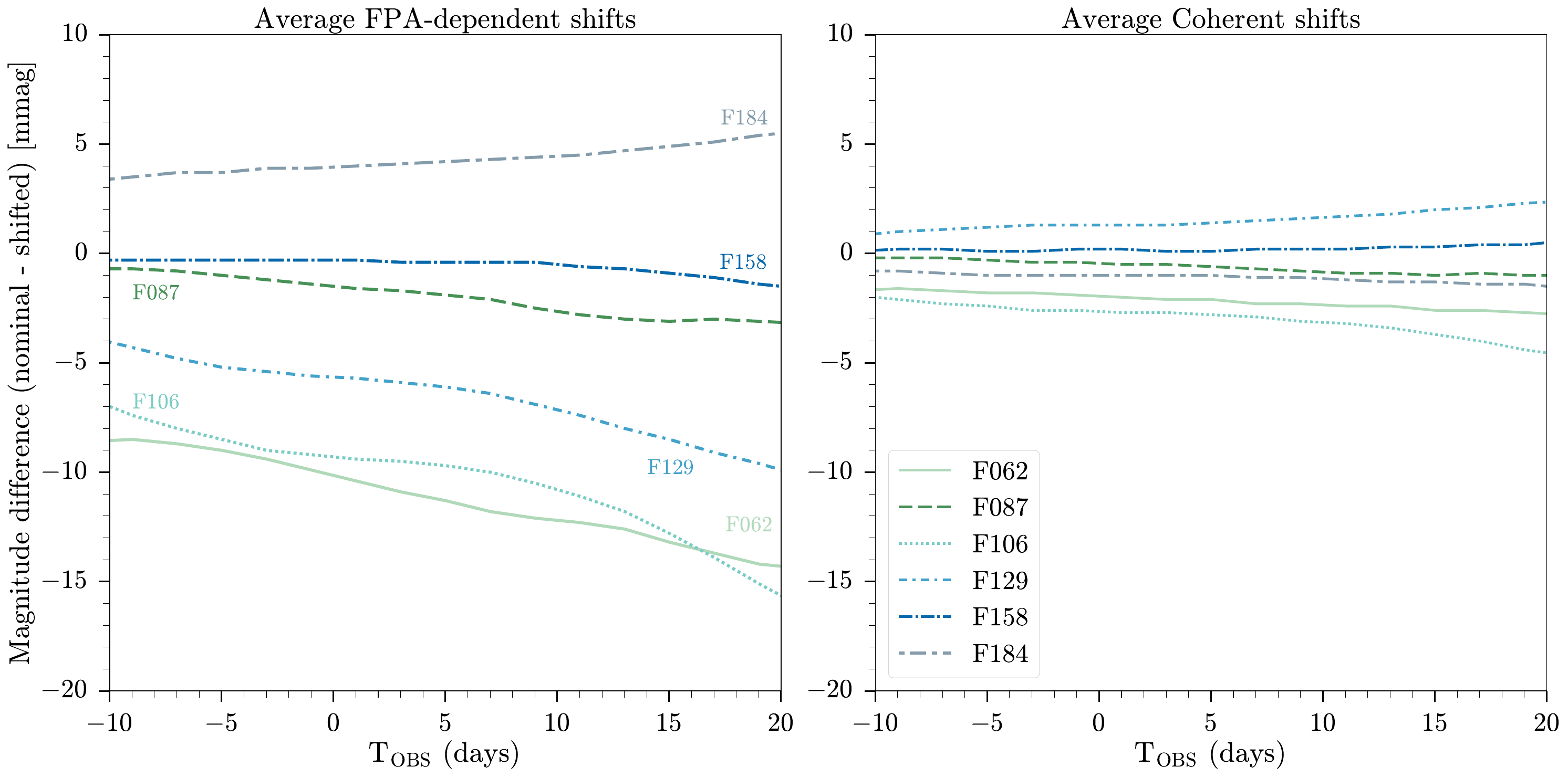}
    \caption{Magnitude difference between nominal and shifted filters for a standard SN Ia at $z=1$. The wavelength shift used here is averaged over all SCAs for each filter. The \textit{x}-axis shows the time of observation of the SN, which is restricted to 10 days before and 20 days after peak brightness. The left panel shows the FPA-dependent shifts and the right panel shows COH-FIXED shifts.}
    \label{fig:mag_template}
\end{figure*}

From our simulations, we obtain light curves for SNe Ia across $0.1 < z < 3$ and falling on each detector on the FPA. For each of these SNe, we calculate the AB magnitude in the respective band defined as \citep{Oke1974}:

\begin{equation}
    M = - 2.5\log_{10}{(\rm flux)} -48.6
\end{equation}

In Figure \ref{fig:mag_template}, we plot the difference in magnitudes calculated using the nominal and shifted filters for a standard SN Ia at $z=1$. We use SCA 2 as our reference SCA and consider all differences relative to SCA 2. In the left panel, we use the mean shift for each filter. In the right panel, we use the coherent shift values listed in Table \ref{tab:wavelengths}. 

For FPA-dependent shifts, F184 is the only filter that exhibits a positive magnitude difference, consistent with the fact that F184 also has the largest effective wavelength shift among all the filters. Because the applied shifts move the filter bandpass toward bluer wavelengths, the magnitude measured with the shifted F184 filter samples a region of the SN Ia SED with higher flux at these epochs, resulting in a brighter inferred magnitude relative to the nominal case---consistent with the positive (nominal - shifted) values seen in the plot for F184.

In the right panel of Figure \ref{fig:mag_template}, we plot the mean of the magnitude difference between nominal and the COH-FIXED shifts. If done with COH-RANDOM shifts instead, we find that the scatter between the realizations is small ($\lesssim 0.1$ mmag), reflecting the tight constraint on absolute wavelength calibration achieved in laboratory testing.

\begin{figure*}[ht!]
    \centering
    \includegraphics[width=2\columnwidth]{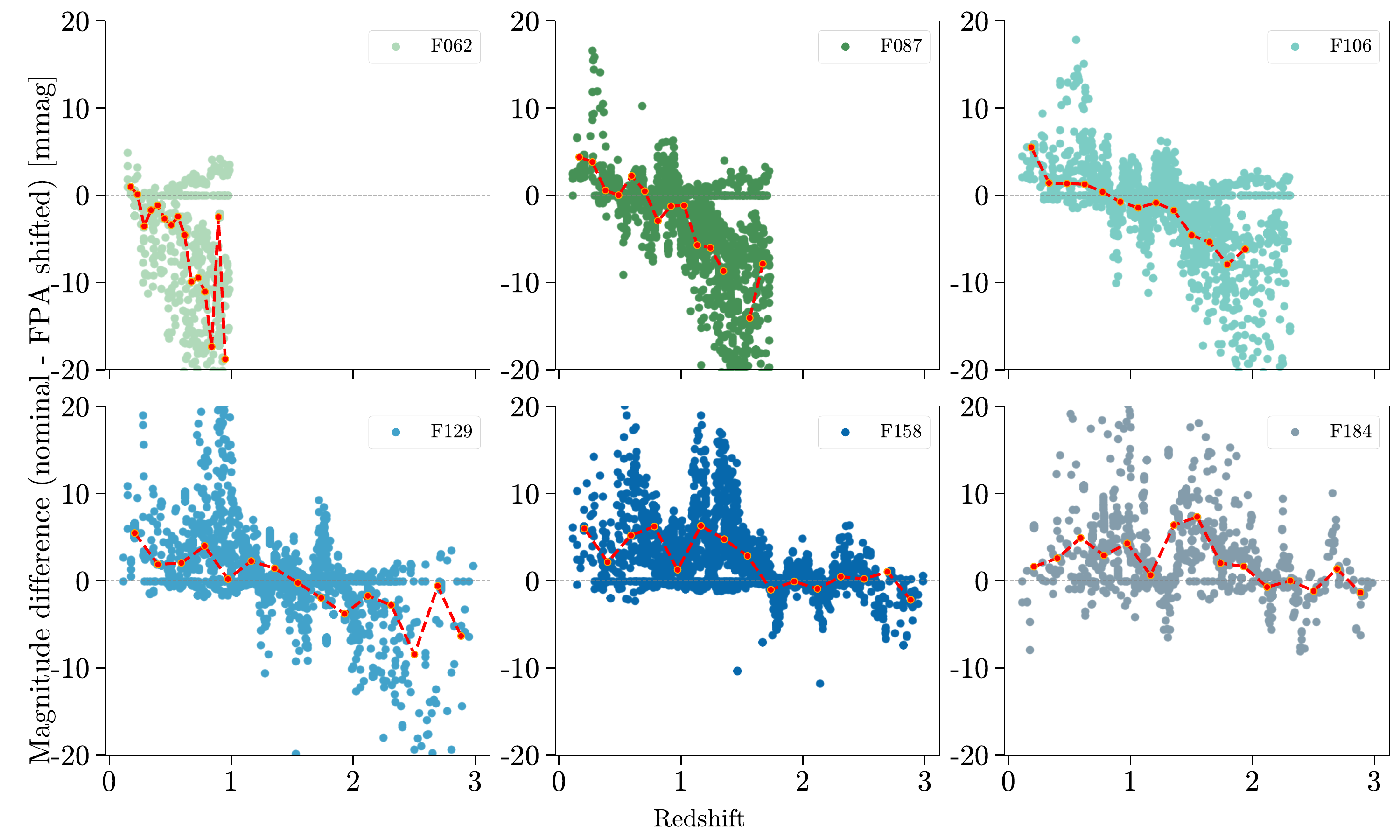}
    \caption{Magnitude difference between nominal and FPA-shifted filters for SNe~Ia observed in each bandpass as a function of redshift. Given the wavelength coverage of each filter, we see that F062 drops off around $z\sim 1$, F087 drops off at $z \sim 1.8$, and F106 at $z\sim 2.4$. The red dashed line in the panels represents the median in each of the 15 redshift bins. We note that the large number of points found at and near $y = 0$ are SNe~Ia detected on the pair of reference SCAs 2 and 11.}
    \label{fig:mag_sim_z}
\end{figure*}

We plot the FPA-dependent chromatic effects as a function of redshift in Figure \ref{fig:mag_sim_z}. One main feature is the systematic dropout of bluer filters at specific redshifts, corresponding to when each filter begins sampling rest-frame wavelengths $\lambda_{\rm rest} \lesssim 3000$ \rm \AA. F062 provides useful photometry until $z \sim 1.0$, beyond which the filters are shifted by $(1+z)$ and samples the UV where SN Ia flux decreases precipitously. Similarly, F087 drops off at $z \sim 1.8$ and F106 at $z \sim 2.4$. These dropout redshifts are precisely where outliers appear in Hubble residuals as SNe have incomplete rest-frame optical coverage.

Within the usable redshift range for each filter, the chromatic effects exhibit distinct evolutionary patterns as seen with the red dashed line representing the median in Figure \ref{fig:mag_sim_z}. F062 shows a mild increase in the magnitude difference between the nominal and shifted filters with redshift. The magnitude difference is $-3.5$ mmag at $z \sim 0.28$, jumping up to -1.16 at $z \sim 0.4$. At $z \sim 1$ where F062 drops out, the magnitude difference is $\sim -18.7$ mmag.

F087 and F106 exhibit some variation as well but are mainly seen to have a decreasing magnitude difference with redshift. Filters F129, F158, and F184 show much more complex behavior, with magnitude differences varying nonmonotonically across their redshift range, reflecting the filter bandpass crossing significant spectral features as redshift increases. This behavior reflects the combined effects of SCA-dependent wavelength variations, intrinsic diversity in SN colors, and light-curve shapes.

\subsection{Nuisance parameters $\alpha$ and $\beta$}

To probe for systematic trends, we calculate the differences in the three components of the Tripp equation ($x_1, c, m_B $). These differences are calculated for each individual SN, allowing us to examine the distribution and redshift evolution of chromatic effects on the standardized distance modulus.

We use the \code{SALT2mu} code \citep[][]{Marriner2011} to determine the nuisance parameters $\alpha$ and $\beta$. The chromatic effects do not significantly change these values: $0.4\%$ for $\alpha$ and $0.17\%$ for $\beta$ for FPA shifts. For coherent shifts, the average change is $\sim 0.07\%$ in $\alpha$ and $\sim 0.2\%$ in $\beta$. Using these values, we calculate the differences from the two samples using the following equations: 

\begin{equation}
    \Delta \alpha x_1 = \alpha_0 x_1 (\rm{nominal}) - \alpha ' x_1 (\rm{shifted})
\end{equation}

\begin{equation}
    \Delta \beta c = \beta_0 c (\rm{nominal}) - \beta ' c (\rm{shifted})
\end{equation}
    
\begin{equation}
     \Delta m_B = m_B (\rm{nominal}) - \textit{m}_B (\rm{shifted})
\end{equation}

$\alpha_0$ and $\beta_0$ are calculated for the nominal dataset with no shifts while $\alpha '$ and $\beta '$ are calculated separately for the shifted datasets.

\begin{figure*}[t!]
    \centering
    \includegraphics[width=2\columnwidth]{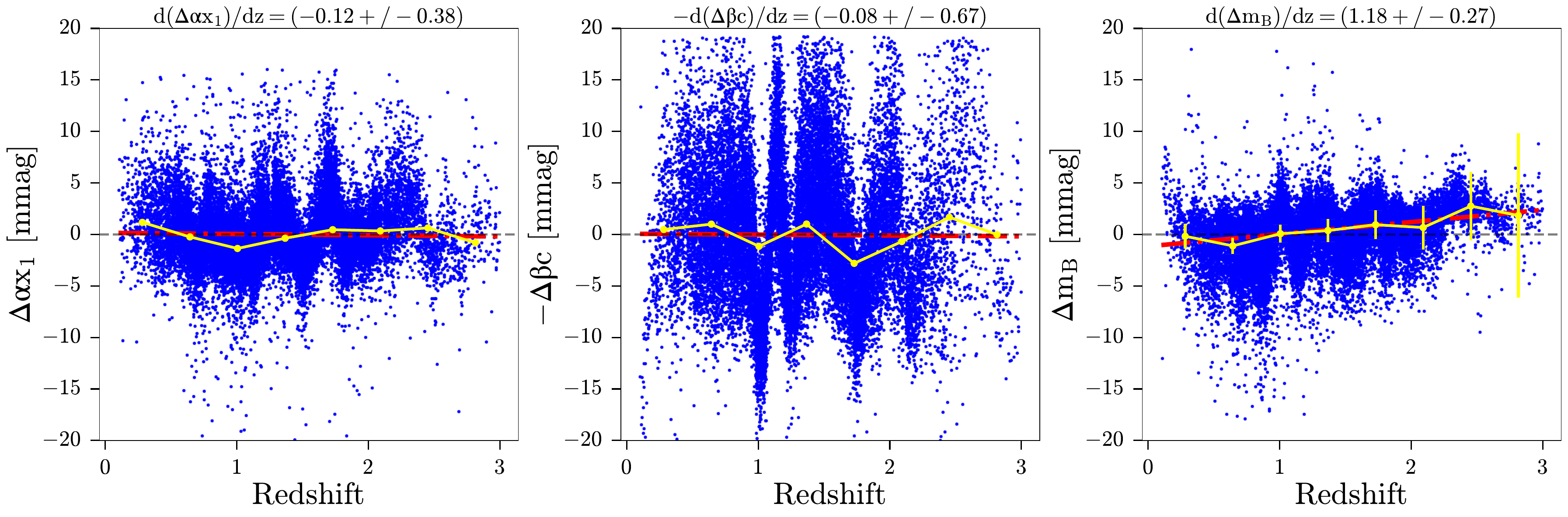}
    \caption{The redshift dependence of $\Delta \alpha x_1$, $\Delta \beta c$, and $\Delta m_B$. Each point is an individual SN from our simulation for FPA-dependent shifts. The yellow solid line connects the median difference for SNe in each redshift bin, and the error bars represent the standard deviation of the impact of the chromatic effects in that bin. Errors in $\Delta \alpha x_1$ and $\Delta \beta c$ are of the order of $< 1$ mmag and hence not starkly visible. The dashed gray line is positioned at $y = 0$ for reference. Finally, the red dashed-dotted line is represents the best linear fit and its slope and uncertainty are given above each panel in mmag. $\Delta \alpha x_1$ and $\Delta \beta c$ have slopes consistent with zero while $\Delta m_B$ has a small slope of 1.18 mmag.}
    \label{fig:ab_fpa_fits}
\end{figure*}

Figure \ref{fig:ab_fpa_fits} shows the redshift dependence of the impact of the chromatic effects on these cosmological parameters for FPA-dependent shifts. We also plot the median values in each redshift bin (in yellow) and fit the best linear approximation (in red). For this linear fit, we exclude outliers beyond $3\sigma$ from the median in each redshift bin to avoid biases from SNe with anomalous photometry or those falling near filter dropout transitions.

The stretch correction term shows no significant redshift dependence with a slope of $\Delta \alpha x_1(z) = (-0.12 \pm 0.38)z$ mmag. In a shot-noise-free analysis, the slope is found to be $(-0.27 \pm 0.43) z$ mmag. The typical magnitude of $\Delta \alpha x_1$ ranges from approximately $-10$ to $+10$ mmag across the full redshift range, with most SNe clustered within $\pm 5$ mmag of zero. Since both the fiducial and shot-noise-free limit estimates are broadly consistent with zero, we place a 95\% upper limit of $|d(\Delta\alpha x_1)/dz| \lesssim 1$ mmag per unit redshift, well below our other chromatic systematics. Any chromatic perturbation to $x_1$ from the reconstruction of rest-frame bands with redshift is therefore not resolved at current precision.

On the other hand, the color correction term exhibits essentially no redshift evolution, with a slope of $-\Delta\beta c(z) = (-0.08 \pm 0.67) z$ mmag. The standard deviation in each redshift bin ranges from 8 to 12 mmag, indicating relatively uniform chromatic effects on color across the SN population. This null result is physically sensible for systematic error analyses. Because our chromatic effects shift all filter bandpasses toward bluer wavelengths by similar fractional amounts, they affect measurements at all wavelengths in a correlated manner. The relative color thus remains largely unchanged, even though the individual magnitudes shift.

The peak magnitude term shows prominent redshift dependence, with a slope of $\Delta m_B(z) = (1.18 \pm 0.27) z$ mmag. This is a significant trend and it indicates that FPA-dependent chromatic effects introduce a significant offset in the inferred rest-frame peak magnitude with a redshift dependence that will manifest as a cosmology systematic.

\begin{figure*}[htbp!]
    \centering
    \includegraphics[width=2\columnwidth]{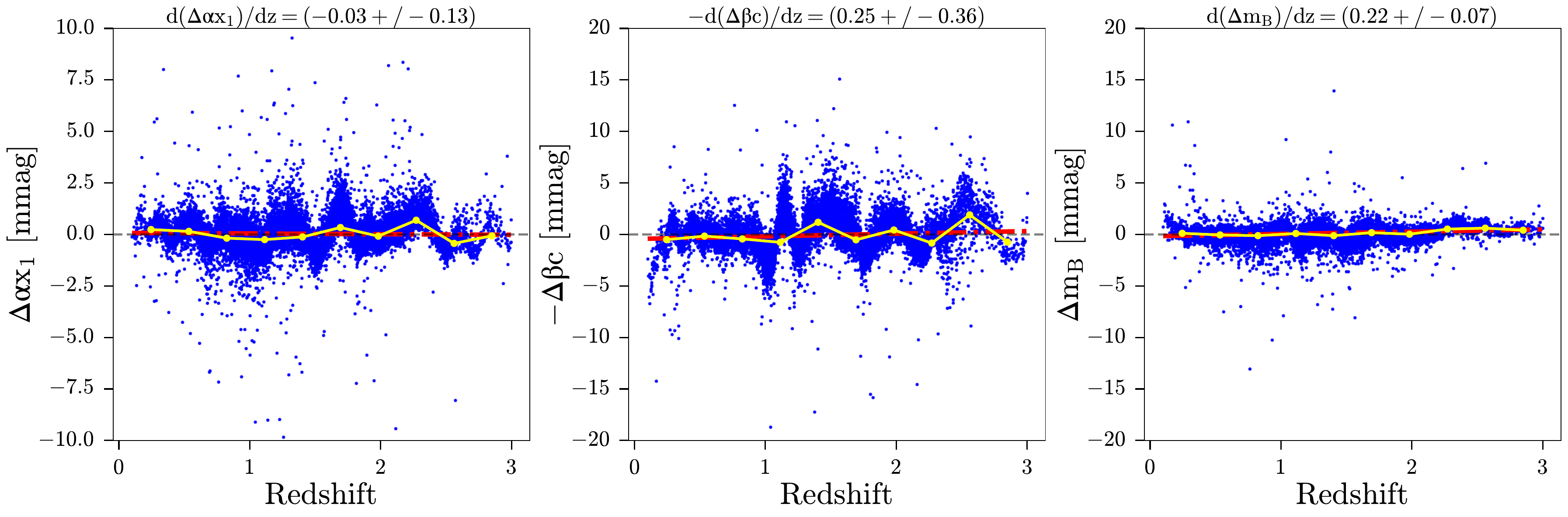}
    \caption{Same as Figure \ref{fig:ab_fpa_fits} but for our simulation with COH-FIXED shifts. $\Delta \alpha x_1$, $\beta c$, and $\Delta m_b$ all have small slopes}.
    \label{fig:ab_coherent_fits}
\end{figure*}

For the COH-FIXED shifts, the slopes for all three Tripp components remain small (Figure \ref{fig:ab_coherent_fits}), with the largest deviation appearing in $\Delta m_B$, whose slope of $(0.22 \pm 0.07) z$ mmag is significantly shallower than the FPA-dependent value of $(1.18 \pm 0.27) z$ mmag. This is consistent with the expectation that a spatially uniform wavelength offset produces little redshift-dependent bias, since all SNe are affected equally regardless of their position on the focal plane or observed redshift.

The statistical uncertainties on our fitted slopes for $\Delta \mu$ provide a natural sensitivity floor: slopes smaller than 1.00 mmag/$z$ (FPA-dependent case) and 0.33 mmag/$z$ (coherent case) would not be distinguishable from zero within our simulation framework, given the sample size of $\sim 22,000$ SNe Ia spanning $0.1 < z < 3$. These limits are set primarily by the intrinsic scatter in SN Ia distances and the finite number of simulated events per redshift bin.

The implications of these results for Roman cosmology are twofold. First, the small overall magnitude of chromatic effects on $\alpha$ and $\beta$ is reassuring, indicating that the fundamental calibration is stable. Second, the mild FPA-dependent redshift dependence in $m_B$ suggest that these effects must be accounted for in cosmological analyses to avoid introducing spurious evolution in the inferred dark energy properties. In the following sections we discuss their impact on cosmology and to what extent the FPA-dependent correction itself must be validated.

\subsection{Cosmology}
\label{sec:cosmo}

We consider the impact of chromatic effects on a range of cosmological models such as a flat-$\Lambda$CDM, flat-$w$CDM, and Flat-$w_0 w_a$CDM. Using the latter of these, we fit for the equation-of-state parameter $w(a) = w_0 + w_a(1 - a)$ which describes the evolution of dark energy with scale factor $a$ \citep[e.g.,][]{Wang2008}. Understanding how systematic wavelength shifts in Roman's filters propagate through to cosmological parameter constraints is crucial for achieving the mission's precision goals in dark energy characterization.

Following \citetalias{Kessler2025}, we include a low-$z$ ($z < 0.09$) set of $\sim 800$ SNe Ia from the proposed LSST WFD baseline strategy. The low-$z$ sample is included to provide more accurate estimates for statistical uncertainties in our cosmology analysis. These SNe are not affected by any filter shifts and are only used to anchor the Hubble diagram. For more details on the low-$z$ sample, we refer the reader to Section 2 of \citetalias{Kessler2025}. We follow the analysis in \citetalias{Kessler2025} (the same as in \S \ref{sec:methods}). This includes light-curve fitting and bias corrections. We impose the same cosmology-grade cuts on our sample as \citetalias{Kessler2025}  which are more conservative than those used in \S \ref{sec:methods} and bring down the number of SNe from $\sim 22,000$ to $\sim 10,000$. These cuts are as follows:

\begin{enumerate}
    \item Each supernova must have at least one observation with $T_{\rm rest} < -5$ days and another with $T_{\rm rest} > +20$ days.
    \item SNe must have SNRSUM $> 40$, where SNRSUM is the sum of SNR observations for $-15 < T_{\rm rest} < +45$ days taken in quadrature.
    \item Each supernova must have at least one observation of SNR $>5$ in three bands.
\end{enumerate}

For each result (FPA-dependent shifts and coherent shifts using COH-RANDOM), we fit the same sample of simulated SNe Ia with the \code{SALT3} light-curve fitter, extract standardized distance moduli using Equation \ref{eq:tripp}, and use \code{wfit} to infer cosmological parameters. The differences in recovered cosmological parameters are denoted by the expressions

\begin{equation}
    \rm \Delta \Omega_{\textit{M}} = \Omega_{\textit{M}}(nominal) - \Omega_{\textit{M}}(shifted)
    \label{eq:dom}
\end{equation}

\begin{equation}
    \rm \Delta {\textit{w}}_{0} = {\textit{w}}_{0}(nominal) - {\textit{w}}_{0}(shifted)
    \label{eq:dw0}
\end{equation}

\begin{equation}
    \rm \Delta {\textit{w}}_{\textit{a}} = {\textit{w}}_{\textit{a}}(nominal) - {\textit{w}}_{\textit{a}}(shifted)
    \label{eq:dwa}
\end{equation}

We use the $w_0w_a$CDM model to constrain cosmological parameters as it provides the most general parameterization of dark energy. The $w_0w_a$CDM model has three free parameters: $\Omega_M$, and the dark energy equation-of-state parameters $w_0$ and $w_a$.  

\begin{deluxetable*}{cccccccc}[]
\centering
\setlength{\tabcolsep}{5pt} % Adjust  horizontal spacing
\renewcommand{\arraystretch}{1.4} % Adjust vertical spacing
\tablehead{
\colhead{Dataset} & \colhead{Prior} & \colhead{$\Delta \Omega_M$} & \colhead{Stat. $\sigma_{\Omega_M}$} & \colhead{$\Delta w_0$} & \colhead{Stat. $\sigma_{w_0}$} & \colhead{$\Delta w_a$} & \colhead{Stat. $\sigma_{w_a}$}}
\startdata
FPA-dependent shifts & CMB prior & -0.003 & 0.004 & -0.066 & 0.025 & 0.236 & 0.114 \\
 & SN-only & -0.005 & 0.013 & -0.066 & 0.03 & 0.270 & 0.317\\ \hline
Half FPA-dependent shifts & CMB prior & -0.001 & 0.004 & -0.032 & 0.025 & 0.116 & 0.114 \\
 & SN-only & -0.003 & 0.013 & -0.033 & 0.03 & 0.138 & 0.317  \\ \hline
Coherent (COH-RANDOM) shifts & CMB prior & 0 & 0.004 & 0.0001 & 0.025 & -0.0004 & 0.114 \\
 & SN-only & 0.001 & 0.013 & 0.0003 & 0.03 & -0.0046 & 0.317 \\ 
\enddata
\title{Systematic differences in cosmological parameters due to chromatic effects}
\caption{The FPA-dependent case includes realistic spatial variation of wavelength shifts across the 18 SCAs, and the case of coherent shifts assumes uniform 0.06\% wavelength offsets across all detectors. Delta values are differences between cosmology obtained from nominal simulations (no filter shifts) and cosmology obtained from shifted simulations (FPA or COH-RANDOM shifts).}
\label{tab:cosmology_table}
\end{deluxetable*}

In our cosmology analysis, we include bias corrections using BBC before constructing a full error covariance matrix. The matrix propagates light-curve fitting uncertainties and enables us to cleanly separate statistical and systematic contributions to the total uncertainty budget. Each realization variant perturbs a single aspect of the light-curve fitting (in this case, the assumed filter transmission), and the resulting $\Delta\mu$ for all SNe in the dataset is used to populate the off-diagonal elements of the covariance matrix. We rebin the Hubble diagram in $x_1$ and $c$ following \citet{Brout2021} and \citet{Kessler_2023}, which reduces the dimensionality of the covariance matrix by grouping SNe~Ia with similar stretch and color properties without the loss of important information needed for systematic `self calibration' and is consistent with the standard BBC methodology used by \citet{Vincenzi2024}, \citet{Popovic2025}, and other modern SN~Ia analyses.

We carry out the cosmology fitting with \code{wfit}, a fast $\chi^2$ minimization code implemented in \code{SNANA}. We conduct two separate cosmology analyses: one incorporating a prior for the cosmic microwave background (CMB), and another using SN data alone.

We fit each data set for cosmology parameters $\Omega_M$, $w_0$, and $w_a$ minimizing as

\begin{equation}
    \chi^2 = \Delta \mu^{T} C_{\rm tot}^{-1} \Delta \mu + \chi^2_{\rm prior}
\end{equation}

where $\chi^2_{\rm prior}$ incorporates a CMB prior on the shift parameter $R \equiv \sqrt{\Omega_M H_0^2}\,\chi(z_*)/c$, where $\chi(z_*)$ is the comoving distance to recombination, with $\sigma_R = 0.0044$ (following \citetalias{Kessler2025}). To avoid bias from the CMB constraint, $R$ is computed from the same cosmology parameters used to generate the simulation.

\begin{figure}[h!]
    \centering
    \includegraphics[width=\columnwidth]{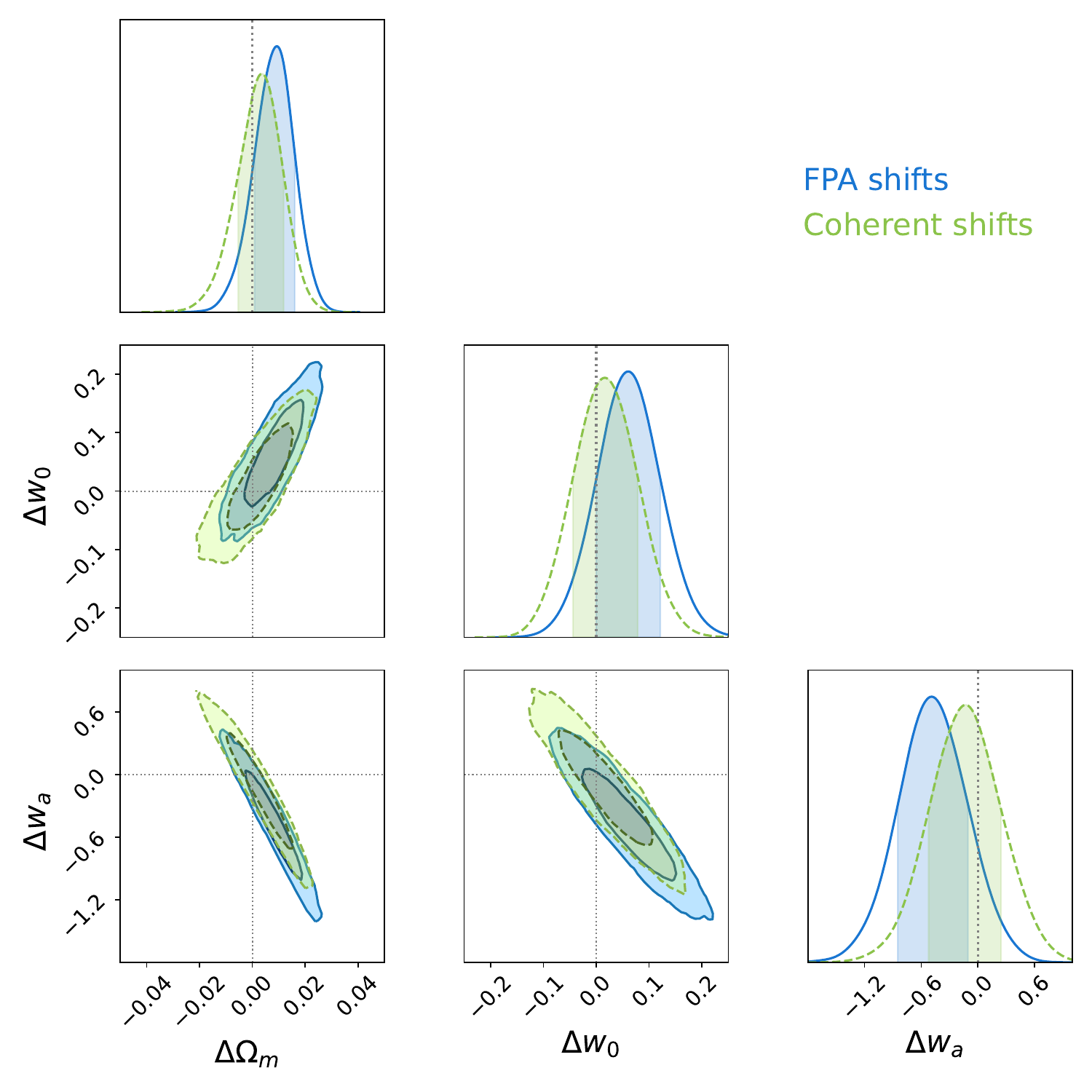}
    \caption{Relative contours of cosmological parameters $\Omega_M$, $w_0$, and $w_a$ for FPA-dependent shifts (blue) and \mbox{COH-RANDOM} shifts (green) as compared to values obtained from the nominal simulation.}
    \label{fig:contours}
\end{figure}

We primarily seek to interpret the relative differences between the shifted and the nominal cases. We report these differences in Table \ref{tab:cosmology_table} from our \code{wfit} outputs along with the statistical uncertainties associated with each parameter. We also create relative contours (relative to cosmology obtained using no filter shifts in the simulation) and present these in Figure \ref{fig:contours}.

Compared to \citetalias{Kessler2025}, our stat-only analysis yields a higher FoM (624 vs.\ 578). This improvement is largely driven by three key differences between our analyses. First, our simulations include a large number of high-$z$ SNe detected in the overlapping WIDE and DEEP fields, which are absent from the \citetalias{Kessler2025} sample. Second, the resulting increase in high-$z$ events gives us smaller $\mu$ errors in this redshift region. Third, because \citetalias{Kessler2025} relies primarily on photo-$z$s, they also have larger $\mu$ errors at lower redshifts.

For FPA-dependent shifts, with a CMB prior, we obtain $\Delta w_0 = -0.066$ with a statistical uncertainty of 0.0253. For $w_a$, we find $\Delta w_a = 0.236$ with $\sigma_{stat} = 0.114$. $\Omega_M$ is well-constrained by the nominal and FPA-shifted datasets, with $\Delta \Omega_M = -0.003$. We obtain an FoM of 590 for the FPA-shifted simulation.

For the case of coherent shifts (COH-RANDOM), we also conduct two separate analyses: one with a CMB prior and one that just uses the SN data. With a CMB prior, the differences are, on average, smaller than the FPA-shifted case for $w_0 w_a$: $\Delta w_0 = 0.0001 , \Delta w_a = -0.0004$. We find $\Omega_M$ of $\Delta \Omega_M = -0.001$.

These findings have important implications for Roman's cosmology. While the systematic biases from chromatic effects are subdominant to expected statistical errors for early analyses, they will become increasingly important as the survey accumulates SNe and statistical uncertainties decrease. The redshift-dependent nature of the chromatic bias is particularly concerning for measurements of $w_a$, which specifically probes dark energy evolution. 

\section{Discussion}
\label{sec:discussion}

\subsection{Stellar photometry}

While our analysis focuses on chromatic effects in the context of SN photometry and cosmology, the wavelength-dependent filter shifts we characterize have broader implications for stellar observations across the focal plane. Complementary work by \citet{Aldoroty2026} has already demonstrated measurable variation in stellar profiles from the OpenUniverse simulations \citep[][]{Troxel2025} through effective modeling of the point-spread function across the FPA. This provides a natural avenue for on-orbit validation and cross calibration of our chromatic effects framework, leveraging stellar observations as an independent and well-understood probe of the same underlying optical systematics.

\begin{figure}[h!]
    \centering
    \includegraphics[width=\columnwidth]{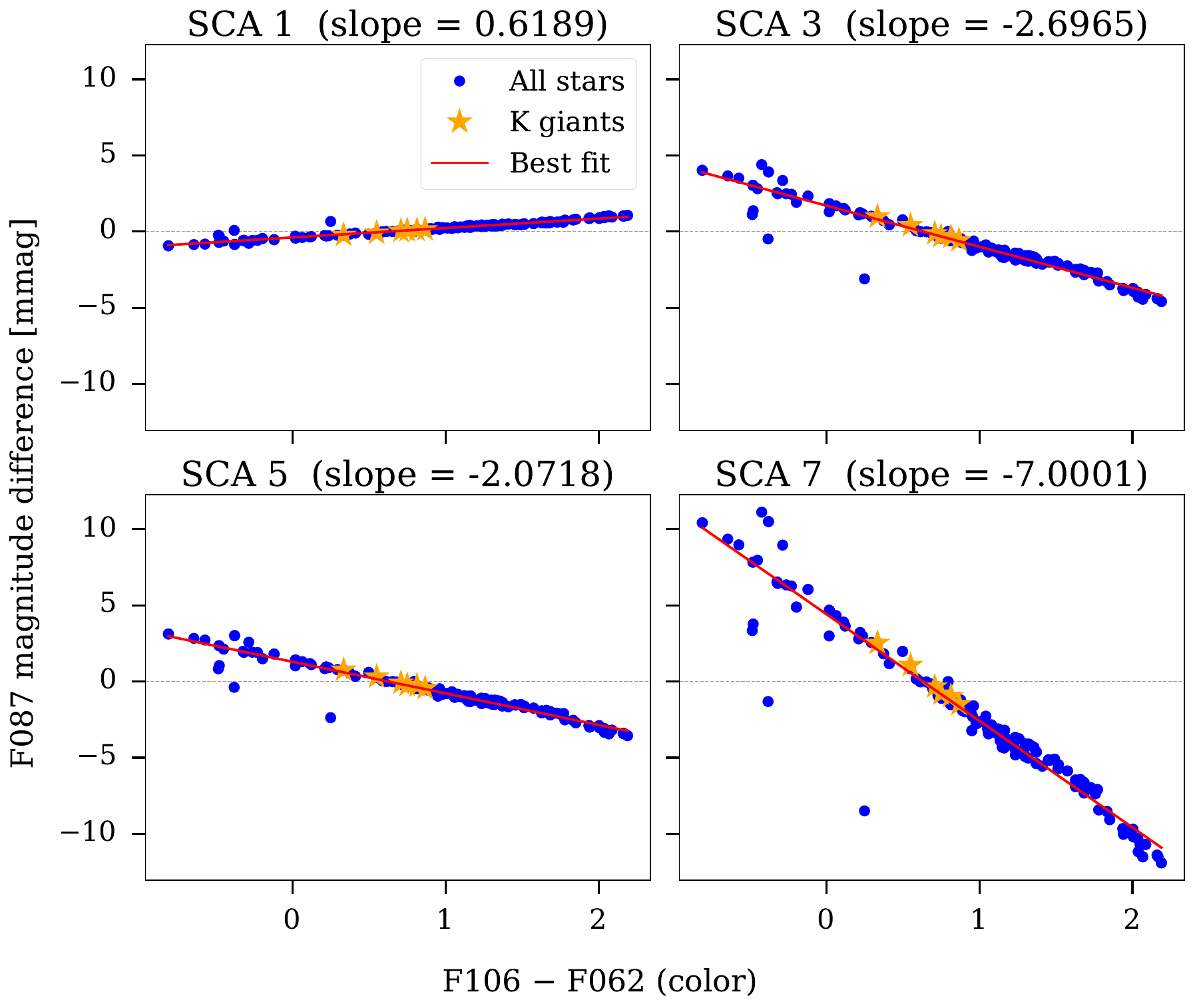}
    \caption{Chromatic effects in F087-band photometry for stars as a function of stellar color (F062 $-$ F106) for four representative SCAs. Each point represents a stellar spectrum from the Pickles library \citep{Pickles1998} spanning spectral types O5 through M7. The magnitude difference on the $y$ axis (nominal $-$ shifted) varies systematically with color due to the wavelength-dependent filter shifts, with redder stars showing a larger impact. Since Roman is expected to be calibrated to K giants in the Large Magellanic Cloud with flight data, we offset the stars in this plot to match such a calibration.}
    \label{fig:stellar_chromatic}
\end{figure}

To demonstrate this, we simulate observations of stars spanning a range of spectral types using the Pickles stellar spectral library \citep{Pickles1998}, which provides empirical flux-calibrated spectra from O-type through M-type stars. We specifically choose this library since the spectra span Roman's wavelength range. Figure \ref{fig:stellar_chromatic} shows the F087-band magnitude difference between nominal and chromatic-corrected photometry as a function of stellar color (F062 $-$ F106) for four representative SCAs. Several key features emerge that make stars ideal probes of chromatic systematics.

First, the chromatic effects in the F087 band varies systematically with stellar color. This trend arises because redder stars have steeper spectral energy distributions, amplifying the photometric impact of wavelength shifts. The approximately linear relationship between color and the chromatic effect provides a direct diagnostic of filter wavelength offsets. Second, different SCAs show distinct chromatic signatures that scale with their respective wavelength shifts. SCA 1 exhibits impacts in magnitude ranging from $\sim -1$ to $+0.6$ mmag and SCA 9 shows the largest range from $-7$ to $+15$ mmag. These differences directly reflect the spatial variations in filter properties across the focal plane that we measure from TVAC testing and incorporate in our SN Ia analysis.

However, unlike SNe, which are transient and observed over limited time baselines, stars provide stable, repeatable photometric targets.  Bright spectrophotometric standard stars observed across the focal plane can achieve relative photometric precision of $\sim 1$ mmag or better \citep{Bohlin2014}, sufficient in principle to detect the chromatic effects of a few millimagnitudes that we predict. But such stars are few in number, and can only probe a limited range of positions; therefore they are primarily useful for a basic confirmation of an existing model.

These results suggest a validation and refinement strategy for Roman's chromatic effects: systematic observations of standard stars distributed across spectral types and observed on all 18 SCAs could directly measure the color-dependent photometric offsets predicted by our measurements of wavelength shift . Specifically, the slope of the chromatic effects versus stellar color (e.g., $\Delta$color in Figure \ref{fig:stellar_chromatic}) is directly proportional to the shift in filter wavelength. By measuring this slope for each SCA and comparing to laboratory-measured filter curves, one can break degeneracies that may be introduced through observations of SNe Ia and even monitor for changes of the SCAs over the missions lifetime. While SNe constrain relative calibration across redshift primarily through the Tripp relation, stars provide absolute wavelength calibration anchored to well-characterized spectral energy distributions. We note that the WFI flat field will be defined so as to ensure that the photometric response of the instrument over the field of view is uniform for a reference spectral energy distribution, which---due to target availability---will most likely be stars in the upper K giant branch; such stars are plentiful in the astrophotometric touchstone field in the Large Magellanic Cloud, which will be the primary target for flat-field definition \citep{Williams2025RomanCalibration}. Any chromatic terms will be therefore be defined differentially with respect to a K giant spectrum, for which we assume there is no residual chromatic zonal term.

Improved chromatic corrections from stellar calibration would directly reduce the systematic uncertainties in SN Ia distance measurements. Currently, our FPA-dependent chromatic effects introduce a slope of $\sim 1.56$ mmag in distance moduli as a function of redshift. If stellar observations refine the measurements of filter wavelength from the current TVAC precision of $\pm 0.5-1 \rm \AA$ to $\pm 0.2$--$0.3$ \AA, the residual chromatic systematic would decrease proportionally. For a Roman SN survey with $\sim 10,000$ cosmology-grade SNe, reducing the per-SN systematic from $0.5$ to $0.2$ mmag would tighten cosmological parameter constraints by $\sim 10-15$\%, particularly for $w_a$, which is most sensitive to redshift-dependent systematics.

\subsection{Cosmology inference and associated limitations}

The distinction between FPA-dependent chromatic effects and coherent residuals has important practical consequences for how Roman's calibration and cosmology pipelines should be designed. The FPA-dependent shifts are fully characterizable from TVAC measurements and can be corrected on a per-object basis once the SCA on which each SN Ia falls is known. The residual coherent term then represents the floor of the calibration uncertainty: the irreducible error that remains if the spectrophotometric response model is imperfect at the 0.06\% level. Our results show that this residual produces $\Delta w_a = -0.0004$, with a statistical uncertainty $\sigma_{w_a} = 0.114$ for the current survey design.

To understand how stringently the FPA-dependent shifts must be characterized, we tested a scenario in which the FPA filter shifts are halved in our simulation (see Table~\ref{tab:cosmology_table}), finding that the resulting bias reduces to $\Delta w_a = 0.116$. The near-linear scaling between the magnitude of the filter shifts and the induced bias in $w_a$ suggests that the relationship is approximately proportional, providing a straightforward way to translate calibration accuracy requirements into cosmological constraints. 

\begin{figure}[h!]
    \centering
    \includegraphics[width=\columnwidth]{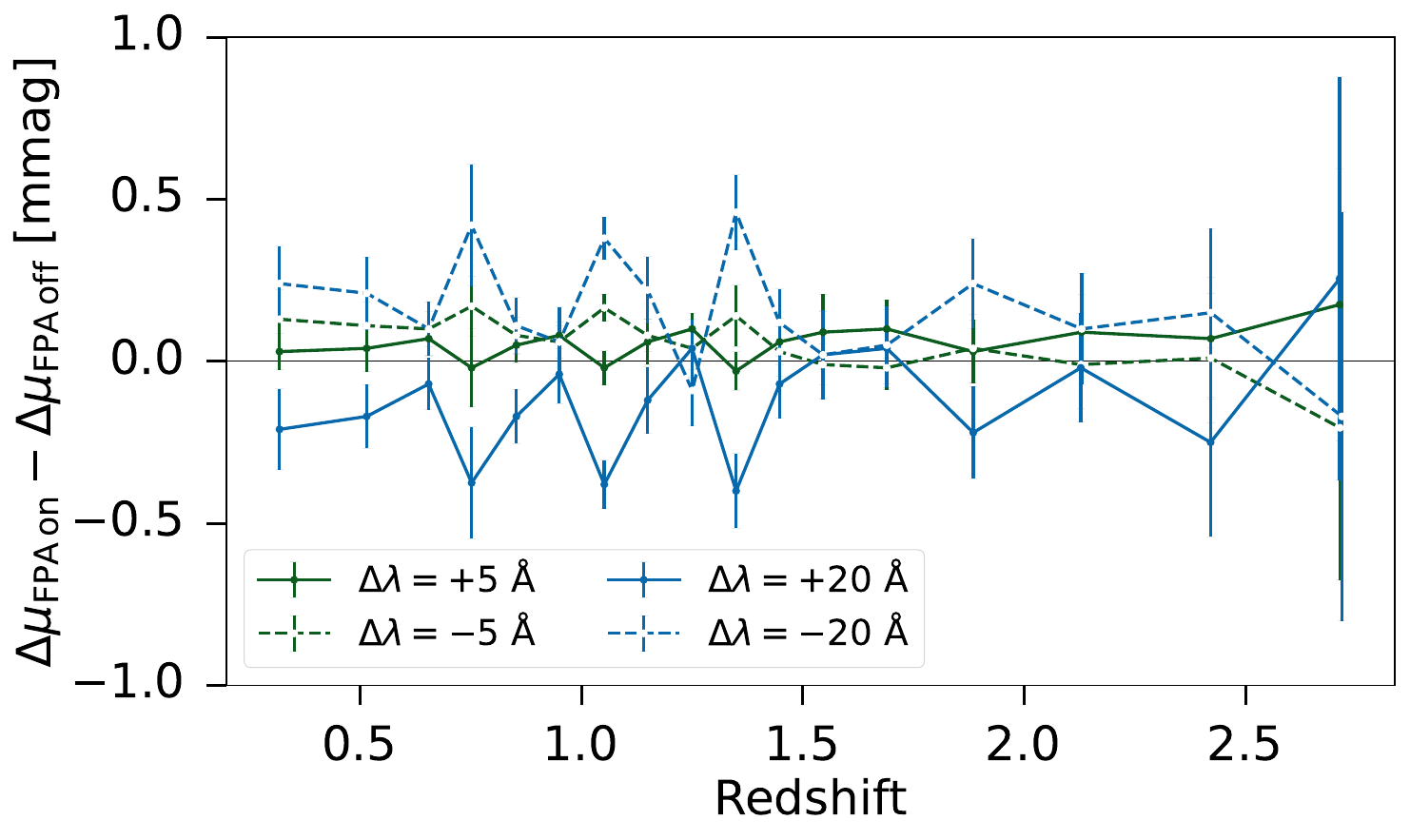}
    \caption{Shifts in Hubble diagram between simulations that include an SED offset with FPA variations ($\Delta\mu_{\rm FPA\,on}$) and those without them ($\Delta\mu_{\rm FPA\,off}$), for $\delta\lambda=\pm5 \rm \AA$ (green) and $\pm20 \rm \AA $ (blue); solid lines are $+\delta\lambda$, dashed lines $-\delta\lambda$. Points show the median in each redshift bin, with error bars giving the uncertainty on the median.}
    \label{fig:sed_bias}
\end{figure}

We note that our study of these chromatic effects assumes that the SALT3 spectral surface represents the ``true'' SN~Ia SED. Limitations in the SALT3 model and our knowledge of the true SN~Ia SED could therefore lead to different corrections than those computed here. To test this, we simulated SNe~Ia with the SALT3 surface shifted in wavelength by $\delta\lambda = \pm5$ and $\pm20 \rm \AA$. A $5 \rm \AA$ shift is comparable to the model--data offsets seen in \citet{Kenworthy2021}. For each $\delta\lambda$, we ran identical (same random seed) simulations with and without the variations in FPA throughput. The difference in $\mu$ between the two simulations isolates any bias that the SED offset could induce in our corrections. The SED offset by itself biases distances by up to $\sim 2$ mmag for $5 \rm \AA$ and $\sim 8$ mmag for $20 \rm \AA$. This term is present without any variations in throughput and is part of the general SALT3 model uncertainty. The additional bias in our corrections is $\lesssim 0.2$ mmag in every redshift bin for $5 \rm \AA$ and $\lesssim 0.5$ mmag for $20 \rm \AA$, about 5\% of the SED mismatch itself. This is shown in Figure \ref{fig:sed_bias}. Our chromatic corrections are therefore robust to plausible wavelength offsets between the true SN~Ia SED and the SALT3 model.

We also note that water ice deposition on the FPA will significantly alter the passband by altering the transmission of the detector's antireflective coatings, introducing additional chromatic terms; the impact of these effects on cosmological constraints is the subject of ongoing study and will be addressed in a future paper.

\section{Summary}
\label{sec:summary}

In this paper, we have presented the first study of the impact of chromatic effects on SNe Ia cosmology simulations designed for the Nancy Grace Roman Space Telescope. Our main results are as follows:

\textit{(1) Single-epoch photometry:} Chromatic effects based on focal plane shifts introduce magnitude shifts of +0.4 to -17.8 mmag depending on the filter, with F184 showing a unique positive shift due to its large wavelength offset ($-39 \rm \AA$) and position at the steep red end of the SN spectrum. The magnitude of these effects varies with SN Ia phase, increasing at late times when nebular emission features become prominent. Shifting the filter bandpass with a coherent 0.06\% wavelength shift results in the need for corrections of the order of +3.3 to -5.3 mmag.

\textit{(2) Redshift dependence:} The dominant systematic we uncover is a linear redshift-dependent bias in distance moduli: $\Delta\mu(z) = (1.56 \pm 1.00) z$ mmag for FPA shifts and $\Delta\mu(z) = (0.60 \pm 0.33)z$ mmag for coherent (COH-FIXED) shifts. This trend arises from the combination of increasingly large wavelength shifts in redder filters and the changing rest-frame wavelength coverage as different observed-frame filters are used to reconstruct rest-frame magnitudes at different redshifts.

\textit{(3) Nuisance parameters:} For FPA-dependent shifts, the systematic uncertainties propagate primarily through the peak magnitude term ($\Delta m_B(z) = (1.18 \pm 0.27)z $ mmag), and contributions from the stretch and color terms are negligible. For coherent shifts, the slopes of all three parameters ($\alpha x_1$, $c$, and $m_B$) are broadly consistent with zero.

\textit{(4) Cosmological parameters:} We fit our data to a $w_0w_a$CDM model and quantify the biases in $\Omega_M$, $w_0$, and $w_a$ induced by filter shifts and wavelength calibration errors. For FPA shifts, we find $\Delta w_0 = -0.066$ and $\Delta w_a = 0.236$; for coherent shifts, $\Delta w_0 = 0.0001$ and $\Delta w_a = -0.0004$.

\textit{(5) Stellar photometry:} We also show that chromatic effects are measurable in stellar photometry, with the impact of the effects ranging from $\sim$1 mmag (SCA 1) to $\sim$20 mmag (SCA 9) depending on stellar color and SCA position. The approximately linear relationship between color and chromatic effects enables direct measurement of filter wavelength shifts from on-orbit stellar observations, providing independent validation of our TVAC-based measurements and the ability to detect any temporal evolution in filter properties. 

Given that $w_a$ is among the most challenging parameters to constrain and that systematics at this level could meaningfully compromise Roman's science return, these chromatic effects cannot be neglected. As statistical uncertainties shrink with larger SNe Ia samples from Roman and LSST, even biases that are currently below $0.5\sigma$ in the $w_0 w_a$ plane will become increasingly significant.

\section*{Data Availability}
The \code{SNANA} and \code{Pippin} configuration files used in this analysis, along with the detector-specific transmission curves, are available here: https://github.com/Roman-Supernova-PIT/chromatic-effects.

\section*{Acknowledgments}

We thank the reviewer for a careful reading of the manuscript and for thoughtful feedback. We thank Susana Deustua for insightful conversations and guidance. D.B. acknowledges support from the Boston University Rafik B. Hariri Institute for Computing and Computational Science \& Engineering through its Junior Faculty Fellows Program. Funding for the Roman Supernova Project Infrastructure Team has been provided by NASA under contract to 80NSSC24M0023. The material is based upon work supported by NASA under award No. 80GSFC24M0006 to R.H. This research used resources of the National Energy Research Scientific Computing Center, which is supported by the Office of Science of the U.S. Department of Energy using award No. HEP-ERCAP32751. This work was carried out at Boston University, which stands on the ancestral lands of the Wampanoag and the Massachusett People. 

\software{\code{astropy} \citep{Astropy2018}, \code{ChainConsumer} \citep{Hinton2016}, \code{matplotlib} \citep{Hunter:2007}, \code{Pippin} \citep{Hinton2020}, \code{SALT2MU} \citep{Marriner2011}, \code{SALT3} \citep{Guy2007, Taylor2023}, \code{SNANA} \citep{Kessler2009}, \code{scipy} \citep{2020SciPy-NMeth}}

\bibliographystyle{aasjournal}
\bibliography{bib}

@ARTICLE{Spergel2013,
       author = {{Spergel}, D. and {Gehrels}, N. and {Breckinridge}, J. and {Donahue}, M. and {Dressler}, A. and {Gaudi}, B.~S. and {Greene}, T. and {Guyon}, O. and {Hirata}, C. and {Kalirai}, J. and {Kasdin}, N.~J. and {Moos}, W. and {Perlmutter}, S. and {Postman}, M. and {Rauscher}, B. and {Rhodes}, J. and {Wang}, Y. and {Weinberg}, D. and {Centrella}, J. and {Traub}, W. and {Baltay}, C. and {Colbert}, J. and {Bennett}, D. and {Kiessling}, A. and {Macintosh}, B. and {Merten}, J. and {Mortonson}, M. and {Penny}, M. and {Rozo}, E. and {Savransky}, D. and {Stapelfeldt}, K. and {Zu}, Y. and {Baker}, C. and {Cheng}, E. and {Content}, D. and {Dooley}, J. and {Foote}, M. and {Goullioud}, R. and {Grady}, K. and {Jackson}, C. and {Kruk}, J. and {Levine}, M. and {Melton}, M. and {Peddie}, C. and {Ruffa}, J. and {Shaklan}, S.},
        title = "{Wide-Field InfraRed Survey Telescope-Astrophysics Focused Telescope Assets WFIRST-AFTA Final Report}",
      journal = {arXiv e-prints},
         year = 2013,
        month = may,
          eid = {arXiv:1305.5422},
        pages = {arXiv:1305.5422},
          doi = {10.48550/arXiv.1305.5422},
archivePrefix = {arXiv},
       eprint = {1305.5422},
 primaryClass = {astro-ph.IM},
       adsurl = {https://ui.adsabs.harvard.edu/abs/2013arXiv1305.5422S}
}

@article{Taylor2021,
   title={A revised SALT2 surface for fitting Type Ia supernova light curves},
   volume={504},
   ISSN={1365-2966},
   url={http://dx.doi.org/10.1093/mnras/stab962},
   DOI={10.1093/mnras/stab962},
   number={3},
   journal={Monthly Notices of the Royal Astronomical Society},
   publisher={Oxford University Press (OUP)},
   author={Taylor, G and Lidman, C and Tucker, B E and Brout, D and Hinton, S R and Kessler, R},
   year={2021},
   month=apr, pages={4111–4122} }

@misc{Vincenzi2024,
      title={The Dark Energy Survey Supernova Program: Cosmological Analysis and Systematic Uncertainties}, 
      author={M. Vincenzi and D. Brout and P. Armstrong and B. Popovic and G. Taylor and M. Acevedo and R. Camilleri and R. Chen and T. M. Davis and S. R. Hinton and L. Kelsey and R. Kessler and J. Lee and C. Lidman and A. Möller and H. Qu and M. Sako and B. Sanchez and D. Scolnic and M. Smith and M. Sullivan and P. Wiseman and J. Asorey and B. A. Bassett and D. Carollo and A. Carr and R. J. Foley and C. Frohmaier and L. Galbany and K. Glazebrook and O. Graur and E. Kovacs and K. Kuehn and U. Malik and R. C. Nichol and B. Rose and B. E. Tucker and M. Toy and D. L. Tucker and F. Yuan and T. M. C. Abbott and M. Aguena and O. Alves and F. Andrade-Oliveira and J. Annis and D. Bacon and K. Bechtol and G. M. Bernstein and D. Brooks and D. L. Burke and A. Carnero Rosell and J. Carretero and F. J. Castander and C. Conselice and L. N. da Costa and M. E. S. Pereira and S. Desai and H. T. Diehl and P. Doel and I. Ferrero and B. Flaugher and D. Friedel and J. Frieman and J. García-Bellido and M. Gatti and G. Giannini and D. Gruen and R. A. Gruendl and D. L. Hollowood and K. Honscheid and D. Huterer and D. J. James and N. Kuropatkin and O. Lahav and S. Lee and H. Lin and J. L. Marshall and J. Mena-Fernández and F. Menanteau and R. Miquel and A. Palmese and A. Pieres and A. A. Plazas Malagón and A. Porredon and A. K. Romer and A. Roodman and E. Sanchez and D. Sanchez Cid and M. Schubnell and I. Sevilla-Noarbe and E. Suchyta and M. E. C. Swanson and G. Tarle and C. To and A. R. Walker and N. Weaverdyck},
      year={2024},
      eprint={2401.02945},
      archivePrefix={arXiv},
      primaryClass={astro-ph.CO},
      url={https://arxiv.org/abs/2401.02945}, 
}

@ARTICLE{Brout2022,
       author = {{Brout}, Dillon and {Taylor}, Georgie and {Scolnic}, Dan and {Wood}, Charlotte M. and {Rose}, Benjamin M. and {Vincenzi}, Maria and {Dwomoh}, Arianna and {Lidman}, Christopher and {Riess}, Adam and {Ali}, Noor and {Qu}, Helen and {Dai}, Mi},
        title = "{The Pantheon+ Analysis: SuperCal-fragilistic Cross Calibration, Retrained SALT2 Light-curve Model, and Calibration Systematic Uncertainty}",
      journal = {\apj},
         year = 2022,
        month = oct,
       volume = {938},
       number = {2},
          eid = {111},
        pages = {111},
          doi = {10.3847/1538-4357/ac8bcc},
archivePrefix = {arXiv},
       eprint = {2112.03864},
 primaryClass = {astro-ph.CO},
       adsurl = {https://ui.adsabs.harvard.edu/abs/2022ApJ...938..111B}
}

@ARTICLE{DES2024,
       author = {{DES Collaboration} and {Abbott}, T.~M.~C. and {Acevedo}, M. and {Aguena}, M. and {Alarcon}, A. and {Allam}, S. and {Alves}, O. and {Amon}, A. and {Andrade-Oliveira}, F. and {Annis}, J. and {Armstrong}, P. and {Asorey}, J. and {Avila}, S. and {Bacon}, D. and {Bassett}, B.~A. and {Bechtol}, K. and {Bernardinelli}, P.~H. and {Bernstein}, G.~M. and {Bertin}, E. and {Blazek}, J. and {Bocquet}, S. and {Brooks}, D. and {Brout}, D. and {Buckley-Geer}, E. and {Burke}, D.~L. and {Camacho}, H. and {Camilleri}, R. and {Campos}, A. and {Carnero Rosell}, A. and {Carollo}, D. and {Carr}, A. and {Carretero}, J. and {Castander}, F.~J. and {Cawthon}, R. and {Chang}, C. and {Chen}, R. and {Choi}, A. and {Conselice}, C. and {Costanzi}, M. and {da Costa}, L.~N. and {Crocce}, M. and {Davis}, T.~M. and {DePoy}, D.~L. and {Desai}, S. and {Diehl}, H.~T. and {Dixon}, M. and {Dodelson}, S. and {Doel}, P. and {Doux}, C. and {Drlica-Wagner}, A. and {Elvin-Poole}, J. and {Everett}, S. and {Ferrero}, I. and {Fert{\'e}}, A. and {Flaugher}, B. and {Foley}, R.~J. and {Fosalba}, P. and {Friedel}, D. and {Frieman}, J. and {Frohmaier}, C. and {Galbany}, L. and {Garc{\'\i}a-Bellido}, J. and {Gatti}, M. and {Gaztanaga}, E. and {Giannini}, G. and {Glazebrook}, K. and {Graur}, O. and {Gruen}, D. and {Gruendl}, R.~A. and {Gutierrez}, G. and {Hartley}, W.~G. and {Herner}, K. and {Hinton}, S.~R. and {Hollowood}, D.~L. and {Honscheid}, K. and {Huterer}, D. and {Jain}, B. and {James}, D.~J. and {Jeffrey}, N. and {Kasai}, E. and {Kelsey}, L. and {Kent}, S. and {Kessler}, R. and {Kim}, A.~G. and {Kirshner}, R.~P. and {Kovacs}, E. and {Kuehn}, K. and {Lahav}, O. and {Lee}, J. and {Lee}, S. and {Lewis}, G.~F. and {Li}, T.~S. and {Lidman}, C. and {Lin}, H. and {Malik}, U. and {Marshall}, J.~L. and {Martini}, P. and {Mena-Fern{\'a}ndez}, J. and {Menanteau}, F. and {Miquel}, R. and {Mohr}, J.~J. and {Mould}, J. and {Muir}, J. and {M{\"o}ller}, A. and {Neilsen}, E. and {Nichol}, R.~C. and {Nugent}, P. and {Ogando}, R.~L.~C. and {Palmese}, A. and {Pan}, Y.-C. and {Paterno}, M. and {Percival}, W.~J. and {Pereira}, M.~E.~S. and {Pieres}, A. and {Malag{\'o}n}, A.~A. Plazas and {Popovic}, B. and {Porredon}, A. and {Prat}, J. and {Qu}, H. and {Raveri}, M. and {Rodr{\'\i}guez-Monroy}, M. and {Romer}, A.~K. and {Roodman}, A. and {Rose}, B. and {Sako}, M. and {Sanchez}, E. and {Sanchez Cid}, D. and {Schubnell}, M. and {Scolnic}, D. and {Sevilla-Noarbe}, I. and {Shah}, P. and {Smith}, J. Allyn. and {Smith}, M. and {Soares-Santos}, M. and {Suchyta}, E. and {Sullivan}, M. and {Suntzeff}, N. and {Swanson}, M.~E.~C. and {S{\'a}nchez}, B.~O. and {Tarle}, G. and {Taylor}, G. and {Thomas}, D. and {To}, C. and {Toy}, M. and {Troxel}, M.~A. and {Tucker}, B.~E. and {Tucker}, D.~L. and {Uddin}, S.~A. and {Vincenzi}, M. and {Walker}, A.~R. and {Weaverdyck}, N. and {Wechsler}, R.~H. and {Weller}, J. and {Wester}, W. and {Wiseman}, P. and {Yamamoto}, M. and {Yuan}, F. and {Zhang}, B. and {Zhang}, Y.},
        title = "{The Dark Energy Survey: Cosmology Results with {\ensuremath{\sim}}1500 New High-redshift Type Ia Supernovae Using the Full 5 yr Data Set}",
      journal = {\apjl},
         year = 2024,
        month = sep,
       volume = {973},
       number = {1},
          eid = {L14},
        pages = {L14},
          doi = {10.3847/2041-8213/ad6f9f},
archivePrefix = {arXiv},
       eprint = {2401.02929},
 primaryClass = {astro-ph.CO},
       adsurl = {https://ui.adsabs.harvard.edu/abs/2024ApJ...973L..14D}
}

@ARTICLE{Rubin2025a,
       author = {{Rubin}, David and {Aldering}, Greg and {Betoule}, Marc and {Fruchter}, Andy and {Huang}, Xiaosheng and {Kim}, Alex G. and {Lidman}, Chris and {Linder}, Eric and {Perlmutter}, Saul and {Ruiz-Lapuente}, Pilar and {Suzuki}, Nao},
        title = "{Union through UNITY: Cosmology with 2000 SNe Using a Unified Bayesian Framework}",
      journal = {\apj},
         year = 2025,
        month = jun,
       volume = {986},
       number = {2},
          eid = {231},
        pages = {231},
          doi = {10.3847/1538-4357/adc0a5},
archivePrefix = {arXiv},
       eprint = {2311.12098},
 primaryClass = {astro-ph.CO},
       adsurl = {https://ui.adsabs.harvard.edu/abs/2025ApJ...986..231R}
}

@ARTICLE{Adame2025,
       author = {{Adame}, A.~G. and {Aguilar}, J. and {Ahlen}, S. and {Alam}, S. and {Alexander}, D.~M. and {Alvarez}, M. and {Alves}, O. and {Anand}, A. and {Andrade}, U. and {Armengaud}, E. and {Avila}, S. and {Aviles}, A. and {Awan}, H. and {Bailey}, S. and {Baltay}, C. and {Bault}, A. and {Behera}, J. and {BenZvi}, S. and {Beutler}, F. and {Bianchi}, D. and {Blake}, C. and {Blum}, R. and {Brieden}, S. and {Brodzeller}, A. and {Brooks}, D. and {Buckley-Geer}, E. and {Burtin}, E. and {Calderon}, R. and {Canning}, R. and {Carnero Rosell}, A. and {Cereskaite}, R. and {Cervantes-Cota}, J.~L. and {Chabanier}, S. and {Chaussidon}, E. and {Chaves-Montero}, J. and {Chen}, S. and {Chen}, X. and {Claybaugh}, T. and {Cole}, S. and {Cuceu}, A. and {Davis}, T.~M. and {Dawson}, K. and {de la Macorra}, A. and {de Mattia}, A. and {Deiosso}, N. and {Dey}, A. and {Dey}, B. and {Ding}, Z. and {Doel}, P. and {Edelstein}, J. and {Eftekharzadeh}, S. and {Eisenstein}, D.~J. and {Elliott}, A. and {Fagrelius}, P. and {Fanning}, K. and {Ferraro}, S. and {Ereza}, J. and {Findlay}, N. and {Flaugher}, B. and {Font-Ribera}, A. and {Forero-S{\'a}nchez}, D. and {Forero-Romero}, J.~E. and {Garcia-Quintero}, C. and {Gazta{\~n}aga}, E. and {Gil-Marn}, H. and {Gontcho a Gontcho}, S. and {Gonzalez-Morales}, A.~X. and {Gonzalez-Perez}, V. and {Gordon}, C. and {Green}, D. and {Gruen}, D. and {Gsponer}, R. and {Gutierrez}, G. and {Guy}, J. and {Hadzhiyska}, B. and {Hahn}, C. and {Hanif}, M.~M.~S. and {Herrera-Alcantar}, H.~K. and {Honscheid}, K. and {Howlett}, C. and {Huterer}, D. and {Ir{\v{s}}i{\v{c}}}, V. and {Ishak}, M. and {Juneau}, S. and {Kara{\c{c}}ayl{\i}}, N.~G. and {Kehoe}, R. and {Kent}, S. and {Kirkby}, D. and {Kong}, H. and {Kremin}, A. and {Krolewski}, A. and {Lai}, Y. and {Lan}, T.-W. and {Landriau}, M. and {Lang}, D. and {Lasker}, J. and {Le Goff}, J.~M. and {Le Guillou}, L. and {Leauthaud}, A. and {Levi}, M.~E. and {Li}, T.~S. and {Linder}, E. and {Lodha}, K. and {Magneville}, C. and {Manera}, M. and {Margala}, D. and {Martini}, P. and {Maus}, M. and {McDonald}, P. and {Medina-Varela}, L. and {Meisner}, A. and {Mena-Fern{\'a}ndez}, J. and {Miquel}, R. and {Moon}, J. and {Moore}, S. and {Moustakas}, J. and {Mueller}, E. and {Mu{\~n}oz-Guti{\'e}rrez}, A. and {Myers}, A.~D. and {Nadathur}, S. and {Napolitano}, L. and {Neveux}, R. and {Newman}, J.~A. and {Nguyen}, N.~M. and {Nie}, J. and {Niz}, G. and {Noriega}, H.~E. and {Padmanabhan}, N. and {Paillas}, E. and {Palanque-Delabrouille}, N. and {Pan}, J. and {Penmetsa}, S. and {Percival}, W.~J. and {Pieri}, M.~M. and {Pinon}, M. and {Poppett}, C. and {Porredon}, A. and {Prada}, F. and {P{\'e}rez-Fern{\'a}ndez}, A. and {P{\'e}rez-R{\`a}fols}, I. and {Rabinowitz}, D. and {Raichoor}, A. and {Ram{\'\i}rez-P{\'e}rez} C. and {Ramirez-Solano}, S. and {Rashkovetskyi}, M. and {Ravoux}, C. and {Rezaie}, M. and {Rich}, J. and {Rocher}, A. and {Rockosi}, C. and {Roe}, N.~A. and {Rosado-Marin}, A. and {Ross}, A.~J. and {Rossi}, G. and {Ruggeri}, R. and {Ruhlmann-Kleider}, V. and {Samushia}, L. and {Sanchez}, E. and {Saulder}, C. and {Schlafly}, E.~F. and {Schlegel}, D. and {Schubnell}, M. and {Seo}, H. and {Sharples}, R. and {Silber}, J. and {Slosar}, A. and {Smith}, A. and {Sprayberry}, D. and {Swanson}, J. and {Tan}, T. and {Tarl{\'e}}, G. and {Trusov}, S. and {Vaisakh}, R. and {Valcin}, D. and {Valdes}, F. and {Vargas-Maga{\~n}a}, M. and {Verde}, L. and {Walther}, M. and {Wang}, B. and {Wang}, M.~S. and {Weaver}, B.~A. and {Weaverdyck}, N. and {Wechsler}, R.~H. and {Weinberg}, D.~H. and {White}, M. and {Wilson}, M.~J. and {Yu}, J. and {Yu}, Y. and {Yuan}, S. and {Y{\`e}che}, C. and {Zaborowski}, E.~A. and {Zarrouk}, P. and {Zhang}, H. and {Zhao}, C. and {Zhao}, R. and {Zhou}, R. and {Zou}, H. and {DESI Collaboration}},
        title = "{DESI 2024 III: baryon acoustic oscillations from galaxies and quasars}",
      journal = {\jcap},
         year = 2025,
        month = apr,
       volume = {2025},
       number = {4},
          eid = {012},
        pages = {012},
          doi = {10.1088/1475-7516/2025/04/012},
archivePrefix = {arXiv},
       eprint = {2404.03000},
 primaryClass = {astro-ph.CO},
       adsurl = {https://ui.adsabs.harvard.edu/abs/2025JCAP...04..012A}
}

@ARTICLE{Spergel2015,
       author = {{Spergel}, D. and {Gehrels}, N. and {Baltay}, C. and {Bennett}, D. and {Breckinridge}, J. and {Donahue}, M. and {Dressler}, A. and {Gaudi}, B.~S. and {Greene}, T. and {Guyon}, O. and {Hirata}, C. and {Kalirai}, J. and {Kasdin}, N.~J. and {Macintosh}, B. and {Moos}, W. and {Perlmutter}, S. and {Postman}, M. and {Rauscher}, B. and {Rhodes}, J. and {Wang}, Y. and {Weinberg}, D. and {Benford}, D. and {Hudson}, M. and {Jeong}, W. -S. and {Mellier}, Y. and {Traub}, W. and {Yamada}, T. and {Capak}, P. and {Colbert}, J. and {Masters}, D. and {Penny}, M. and {Savransky}, D. and {Stern}, D. and {Zimmerman}, N. and {Barry}, R. and {Bartusek}, L. and {Carpenter}, K. and {Cheng}, E. and {Content}, D. and {Dekens}, F. and {Demers}, R. and {Grady}, K. and {Jackson}, C. and {Kuan}, G. and {Kruk}, J. and {Melton}, M. and {Nemati}, B. and {Parvin}, B. and {Poberezhskiy}, I. and {Peddie}, C. and {Ruffa}, J. and {Wallace}, J.~K. and {Whipple}, A. and {Wollack}, E. and {Zhao}, F.},
        title = "{Wide-Field InfrarRed Survey Telescope-Astrophysics Focused Telescope Assets WFIRST-AFTA 2015 Report}",
      journal = {arXiv e-prints},
         year = 2015,
        month = mar,
          eid = {arXiv:1503.03757},
        pages = {arXiv:1503.03757},
          doi = {10.48550/arXiv.1503.03757},
archivePrefix = {arXiv},
       eprint = {1503.03757},
 primaryClass = {astro-ph.IM},
       adsurl = {https://ui.adsabs.harvard.edu/abs/2015arXiv150303757S}
}

@ARTICLE{Akeson2019,
       author = {{Akeson}, Rachel and {Armus}, Lee and {Bachelet}, Etienne and {Bailey}, Vanessa and {Bartusek}, Lisa and {Bellini}, Andrea and {Benford}, Dominic and {Bennett}, David and {Bhattacharya}, Aparna and {Bohlin}, Ralph and {Boyer}, Martha and {Bozza}, Valerio and {Bryden}, Geoffrey and {Calchi Novati}, Sebastiano and {Carpenter}, Kenneth and {Casertano}, Stefano and {Choi}, Ami and {Content}, David and {Dayal}, Pratika and {Dressler}, Alan and {Dor{\'e}}, Olivier and {Fall}, S. Michael and {Fan}, Xiaohui and {Fang}, Xiao and {Filippenko}, Alexei and {Finkelstein}, Steven and {Foley}, Ryan and {Furlanetto}, Steven and {Kalirai}, Jason and {Gaudi}, B. Scott and {Gilbert}, Karoline and {Girard}, Julien and {Grady}, Kevin and {Greene}, Jenny and {Guhathakurta}, Puragra and {Heinrich}, Chen and {Hemmati}, Shoubaneh and {Hendel}, David and {Henderson}, Calen and {Henning}, Thomas and {Hirata}, Christopher and {Ho}, Shirley and {Huff}, Eric and {Hutter}, Anne and {Jansen}, Rolf and {Jha}, Saurabh and {Johnson}, Samson and {Jones}, David and {Kasdin}, Jeremy and {Kelly}, Patrick and {Kirshner}, Robert and {Koekemoer}, Anton and {Kruk}, Jeffrey and {Lewis}, Nikole and {Macintosh}, Bruce and {Madau}, Piero and {Malhotra}, Sangeeta and {Mandel}, Kaisey and {Massara}, Elena and {Masters}, Daniel and {McEnery}, Julie and {McQuinn}, Kristen and {Melchior}, Peter and {Melton}, Mark and {Mennesson}, Bertrand and {Peeples}, Molly and {Penny}, Matthew and {Perlmutter}, Saul and {Pisani}, Alice and {Plazas}, Andr{\'e}s and {Poleski}, Radek and {Postman}, Marc and {Ranc}, Cl{\'e}ment and {Rauscher}, Bernard and {Rest}, Armin and {Roberge}, Aki and {Robertson}, Brant and {Rodney}, Steven and {Rhoads}, James and {Rhodes}, Jason and {Ryan}, Jr., Russell and {Sahu}, Kailash and {Sand}, David and {Scolnic}, Dan and {Seth}, Anil and {Shvartzvald}, Yossi and {Siellez}, Karelle and {Smith}, Arfon and {Spergel}, David and {Stassun}, Keivan and {Street}, Rachel and {Strolger}, Louis-Gregory and {Szalay}, Alexander and {Trauger}, John and {Troxel}, M.~A. and {Turnbull}, Margaret and {van der Marel}, Roeland and {von der Linden}, Anja and {Wang}, Yun and {Weinberg}, David and {Williams}, Benjamin and {Windhorst}, Rogier and {Wollack}, Edward and {Wu}, Hao-Yi and {Yee}, Jennifer and {Zimmerman}, Neil},
        title = "{The Wide Field Infrared Survey Telescope: 100 Hubbles for the 2020s}",
      journal = {arXiv e-prints},
         year = 2019,
        month = feb,
          eid = {arXiv:1902.05569},
        pages = {arXiv:1902.05569},
          doi = {10.48550/arXiv.1902.05569},
archivePrefix = {arXiv},
       eprint = {1902.05569},
 primaryClass = {astro-ph.IM},
       adsurl = {https://ui.adsabs.harvard.edu/abs/2019arXiv190205569A}
}

@ARTICLE{Hounsell2018,
       author = {{Hounsell}, R. and {Scolnic}, D. and {Foley}, R.~J. and {Kessler}, R. and {Miranda}, V. and {Avelino}, A. and {Bohlin}, R.~C. and {Filippenko}, A.~V. and {Frieman}, J. and {Jha}, S.~W. and {Kelly}, P.~L. and {Kirshner}, R.~P. and {Mandel}, K. and {Rest}, A. and {Riess}, A.~G. and {Rodney}, S.~A. and {Strolger}, L.},
        title = "{Simulations of the WFIRST Supernova Survey and Forecasts of Cosmological Constraints}",
      journal = {\apj},
         year = 2018,
        month = nov,
       volume = {867},
       number = {1},
          eid = {23},
        pages = {23},
          doi = {10.3847/1538-4357/aac08b},
archivePrefix = {arXiv},
       eprint = {1702.01747},
 primaryClass = {astro-ph.IM},
       adsurl = {https://ui.adsabs.harvard.edu/abs/2018ApJ...867...23H}
}

@ARTICLE{Rose2021,
       author = {{Rose}, B.~M. and {Baltay}, C. and {Hounsell}, R. and {Macias}, P. and {Rubin}, D. and {Scolnic}, D. and {Aldering}, G. and {Bohlin}, R. and {Dai}, M. and {Deustua}, S.~E. and {Foley}, R.~J. and {Fruchter}, A. and {Galbany}, L. and {Jha}, S.~W. and {Jones}, D.~O. and {Joshi}, B.~A. and {Kelly}, P.~L. and {Kessler}, R. and {Kirshner}, R.~P. and {Mandel}, K.~S. and {Perlmutter}, S. and {Pierel}, J. and {Qu}, H. and {Rabinowitz}, D. and {Rest}, A. and {Riess}, A.~G. and {Rodney}, S. and {Sako}, M. and {Siebert}, M.~R. and {Strolger}, L. and {Suzuki}, N. and {Thorp}, S. and {Van Dyk}, S.~D. and {Wang}, K. and {Ward}, S.~M. and {Wood-Vasey}, W.~M.},
        title = "{A Reference Survey for Supernova Cosmology with the Nancy Grace Roman Space Telescope}",
      journal = {arXiv e-prints},
         year = 2021,
        month = nov,
          eid = {arXiv:2111.03081},
        pages = {arXiv:2111.03081},
          doi = {10.48550/arXiv.2111.03081},
archivePrefix = {arXiv},
       eprint = {2111.03081},
 primaryClass = {astro-ph.CO},
       adsurl = {https://ui.adsabs.harvard.edu/abs/2021arXiv211103081R}
}

@ARTICLE{Rose2025,
       author = {{Rose}, B.~M. and {Vincenzi}, M. and {Hounsell}, R. and {Qu}, H. and {Aldoroty}, L. and {Scolnic}, D. and {Kessler}, R. and {Macias}, P. and {Brout}, D. and {Acevedo}, M. and {Chen}, R.~C. and {Gomez}, S. and {Peterson}, E. and {Rubin}, D. and {Sako}, M. and {the Roman Supernova Project Infrastructure Team}},
        title = "{The Hourglass Simulation: A Catalog for the Roman High-latitude Time-domain Core Community Survey}",
      journal = {\apj},
         year = 2025,
        month = jul,
       volume = {988},
       number = {1},
          eid = {65},
        pages = {65},
          doi = {10.3847/1538-4357/ade1d6},
archivePrefix = {arXiv},
       eprint = {2506.05161},
 primaryClass = {astro-ph.IM},
       adsurl = {https://ui.adsabs.harvard.edu/abs/2025ApJ...988...65R}
}

@ARTICLE{Kessler2025,
       author = {{Kessler}, Richard and {Hounsell}, Rebekah and {Joshi}, Bhavin and {Rubin}, David and {Sako}, Masao and {Chen}, Rebecca and {Miranda}, Vivian and {Rose}, Benjamin. M.},
        title = "{Cosmology Constraints from Type Ia Supernova Simulations of the Nancy Grace Roman Space Telescope Strategy Recommended by the High Latitude Time Domain Survey Definition Committee}",
      journal = {arXiv e-prints},
         year = 2025,
        month = jun,
          eid = {arXiv:2506.04402},
        pages = {arXiv:2506.04402},
          doi = {10.48550/arXiv.2506.04402},
archivePrefix = {arXiv},
       eprint = {2506.04402},
 primaryClass = {astro-ph.CO},
       adsurl = {https://ui.adsabs.harvard.edu/abs/2025arXiv250604402K}
}

@article{Kessler_2023,
   title={Binning is Sinning: Redemption for Hubble Diagram Using Photometrically Classified Type Ia Supernovae},
   volume={952},
   ISSN={2041-8213},
   url={http://dx.doi.org/10.3847/2041-8213/ace34d},
   DOI={10.3847/2041-8213/ace34d},
   number={1},
   journal={The Astrophysical Journal Letters},
   publisher={American Astronomical Society},
   author={Kessler, R. and Vincenzi, M. and Armstrong, P.},
   year={2023},
   month=jul, pages={L8} }

@ARTICLE{Betoule2014,
       author = {{Betoule}, M. and {Kessler}, R. and {Guy}, J. and {Mosher}, J. and {Hardin}, D. and {Biswas}, R. and {Astier}, P. and {El-Hage}, P. and {Konig}, M. and {Kuhlmann}, S. and {Marriner}, J. and {Pain}, R. and {Regnault}, N. and {Balland}, C. and {Bassett}, B.~A. and {Brown}, P.~J. and {Campbell}, H. and {Carlberg}, R.~G. and {Cellier-Holzem}, F. and {Cinabro}, D. and {Conley}, A. and {D'Andrea}, C.~B. and {DePoy}, D.~L. and {Doi}, M. and {Ellis}, R.~S. and {Fabbro}, S. and {Filippenko}, A.~V. and {Foley}, R.~J. and {Frieman}, J.~A. and {Fouchez}, D. and {Galbany}, L. and {Goobar}, A. and {Gupta}, R.~R. and {Hill}, G.~J. and {Hlozek}, R. and {Hogan}, C.~J. and {Hook}, I.~M. and {Howell}, D.~A. and {Jha}, S.~W. and {Le Guillou}, L. and {Leloudas}, G. and {Lidman}, C. and {Marshall}, J.~L. and {M{\"o}ller}, A. and {Mour{\~a}o}, A.~M. and {Neveu}, J. and {Nichol}, R. and {Olmstead}, M.~D. and {Palanque-Delabrouille}, N. and {Perlmutter}, S. and {Prieto}, J.~L. and {Pritchet}, C.~J. and {Richmond}, M. and {Riess}, A.~G. and {Ruhlmann-Kleider}, V. and {Sako}, M. and {Schahmaneche}, K. and {Schneider}, D.~P. and {Smith}, M. and {Sollerman}, J. and {Sullivan}, M. and {Walton}, N.~A. and {Wheeler}, C.~J.},
        title = "{Improved cosmological constraints from a joint analysis of the SDSS-II and SNLS supernova samples}",
      journal = {\aap},
         year = 2014,
        month = aug,
       volume = {568},
          eid = {A22},
        pages = {A22},
          doi = {10.1051/0004-6361/201423413},
archivePrefix = {arXiv},
       eprint = {1401.4064},
 primaryClass = {astro-ph.CO},
       adsurl = {https://ui.adsabs.harvard.edu/abs/2014A&A...568A..22B}
}

@ARTICLE{Scolnic2018,
       author = {{Scolnic}, D.~M. and {Jones}, D.~O. and {Rest}, A. and {Pan}, Y.~C. and {Chornock}, R. and {Foley}, R.~J. and {Huber}, M.~E. and {Kessler}, R. and {Narayan}, G. and {Riess}, A.~G. and {Rodney}, S. and {Berger}, E. and {Brout}, D.~J. and {Challis}, P.~J. and {Drout}, M. and {Finkbeiner}, D. and {Lunnan}, R. and {Kirshner}, R.~P. and {Sanders}, N.~E. and {Schlafly}, E. and {Smartt}, S. and {Stubbs}, C.~W. and {Tonry}, J. and {Wood-Vasey}, W.~M. and {Foley}, M. and {Hand}, J. and {Johnson}, E. and {Burgett}, W.~S. and {Chambers}, K.~C. and {Draper}, P.~W. and {Hodapp}, K.~W. and {Kaiser}, N. and {Kudritzki}, R.~P. and {Magnier}, E.~A. and {Metcalfe}, N. and {Bresolin}, F. and {Gall}, E. and {Kotak}, R. and {McCrum}, M. and {Smith}, K.~W.},
        title = "{The Complete Light-curve Sample of Spectroscopically Confirmed SNe Ia from Pan-STARRS1 and Cosmological Constraints from the Combined Pantheon Sample}",
      journal = {\apj},
         year = 2018,
        month = jun,
       volume = {859},
       number = {2},
          eid = {101},
        pages = {101},
          doi = {10.3847/1538-4357/aab9bb},
archivePrefix = {arXiv},
       eprint = {1710.00845},
 primaryClass = {astro-ph.CO},
       adsurl = {https://ui.adsabs.harvard.edu/abs/2018ApJ...859..101S}
}

@ARTICLE{Jones2019,
       author = {{Jones}, D.~O. and {Scolnic}, D.~M. and {Foley}, R.~J. and {Rest}, A. and {Kessler}, R. and {Challis}, P.~M. and {Chambers}, K.~C. and {Coulter}, D.~A. and {Dettman}, K.~G. and {Foley}, M.~M. and {Huber}, M.~E. and {Jha}, S.~W. and {Johnson}, E. and {Kilpatrick}, C.~D. and {Kirshner}, R.~P. and {Manuel}, J. and {Narayan}, G. and {Pan}, Y. -C. and {Riess}, A.~G. and {Schultz}, A.~S.~B. and {Siebert}, M.~R. and {Berger}, E. and {Chornock}, R. and {Flewelling}, H. and {Magnier}, E.~A. and {Smartt}, S.~J. and {Smith}, K.~W. and {Wainscoat}, R.~J. and {Waters}, C. and {Willman}, M.},
        title = "{The Foundation Supernova Survey: Measuring Cosmological Parameters with Supernovae from a Single Telescope}",
      journal = {\apj},
         year = 2019,
        month = aug,
       volume = {881},
       number = {1},
          eid = {19},
        pages = {19},
          doi = {10.3847/1538-4357/ab2bec},
archivePrefix = {arXiv},
       eprint = {1811.09286},
 primaryClass = {astro-ph.CO},
       adsurl = {https://ui.adsabs.harvard.edu/abs/2019ApJ...881...19J}
}

@ARTICLE{Lasker2019,
       author = {{Lasker}, J. and {Kessler}, R. and {Scolnic}, D. and {Brout}, D. and {Burke}, D.~L. and {D'Andrea}, C.~B. and {Davis}, T.~M. and {Hinton}, S.~R. and {Kim}, A.~G. and {Li}, T.~S. and {Lidman}, C. and {Macaulay}, E. and {M{\"o}ller}, A. and {Rykoff}, E.~S. and {Sako}, M. and {Smith}, M. and {Sullivan}, M. and {Swann}, E. and {Tucker}, B.~E. and {Wester}, W. and {Bassett}, B.~A. and {Abbott}, T.~M.~C. and {Allam}, S. and {Annis}, J. and {Avila}, S. and {Bechtol}, K. and {Bertin}, E. and {Brooks}, D. and {Carnero Rosell}, A. and {Carrasco Kind}, M. and {Carretero}, J. and {Castander}, F.~J. and {Calcino}, J. and {Carollo}, D. and {da Costa}, L.~N. and {Davis}, C. and {De Vicente}, J. and {Diehl}, H.~T. and {Doel}, P. and {Drlica-Wagner}, A. and {Flaugher}, B. and {Frieman}, J. and {Garc{\'\i}a-Bellido}, J. and {Gaztanaga}, E. and {Gruen}, D. and {Gruendl}, R.~A. and {Gschwend}, J. and {Gutierrez}, G. and {Hollowood}, D.~L. and {Honscheid}, K. and {Hoormann}, J.~K. and {James}, D.~J. and {Kent}, S. and {Krause}, E. and {Kron}, R. and {Kuehn}, K. and {Kuropatkin}, N. and {Lima}, M. and {Maia}, M.~A.~G. and {Marshall}, J.~L. and {Martini}, P. and {Menanteau}, F. and {Miller}, C.~J. and {Miquel}, R. and {Plazas}, A.~A. and {Sanchez}, E. and {Scarpine}, V. and {Sevilla-Noarbe}, I. and {Smith}, R.~C. and {Soares-Santos}, M. and {Sobreira}, F. and {Suchyta}, E. and {Swanson}, M.~E.~C. and {Tarle}, G. and {Tucker}, D.~L. and {Walker}, A.~R. and {DES Collaboration}},
        title = "{First cosmology results using Type IA supernovae from the dark energy survey: effects of chromatic corrections to supernova photometry on measurements of cosmological parameters}",
      journal = {\mnras},
         year = 2019,
        month = jun,
       volume = {485},
       number = {4},
        pages = {5329-5344},
          doi = {10.1093/mnras/stz619},
archivePrefix = {arXiv},
       eprint = {1811.02380},
 primaryClass = {astro-ph.CO},
       adsurl = {https://ui.adsabs.harvard.edu/abs/2019MNRAS.485.5329L}
}

@ARTICLE{Guy2007,
       author = {{Guy}, J. and {Astier}, P. and {Baumont}, S. and {Hardin}, D. and {Pain}, R. and {Regnault}, N. and {Basa}, S. and {Carlberg}, R.~G. and {Conley}, A. and {Fabbro}, S. and {Fouchez}, D. and {Hook}, I.~M. and {Howell}, D.~A. and {Perrett}, K. and {Pritchet}, C.~J. and {Rich}, J. and {Sullivan}, M. and {Antilogus}, P. and {Aubourg}, E. and {Bazin}, G. and {Bronder}, J. and {Filiol}, M. and {Palanque-Delabrouille}, N. and {Ripoche}, P. and {Ruhlmann-Kleider}, V.},
        title = "{SALT2: using distant supernovae to improve the use of type Ia supernovae as distance indicators}",
      journal = {\aap},
         year = 2007,
        month = apr,
       volume = {466},
       number = {1},
        pages = {11-21},
          doi = {10.1051/0004-6361:20066930},
archivePrefix = {arXiv},
       eprint = {astro-ph/0701828},
 primaryClass = {astro-ph},
       adsurl = {https://ui.adsabs.harvard.edu/abs/2007A&A...466...11G}
}

@ARTICLE{Tripp1998,
       author = {{Tripp}, Robert},
        title = "{A two-parameter luminosity correction for Type IA supernovae}",
      journal = {\aap},
         year = 1998,
        month = mar,
       volume = {331},
        pages = {815-820},
       adsurl = {https://ui.adsabs.harvard.edu/abs/1998A&A...331..815T}
}

@ARTICLE{Lee2023,
       author = {{Lee}, J. and {Acevedo}, M. and {Sako}, M. and {Vincenzi}, M. and {Brout}, D. and {Sanchez}, B. and {Chen}, R. and {Davis}, T.~M. and {Jarvis}, M. and {Scolnic}, D. and {Qu}, H. and {Galbany}, L. and {Kessler}, R. and {Lasker}, J. and {Sullivan}, M. and {Wiseman}, P. and {Aguena}, M. and {Allam}, S. and {Alves}, O. and {Andrade-Oliveira}, F. and {Bertin}, E. and {Bocquet}, S. and {Brooks}, D. and {Burke}, D.~L. and {Carnero Rosell}, A. and {Carrasco Kind}, M. and {Carretero}, J. and {Costanzi}, M. and {da Costa}, L.~N. and {Pereira}, M.~E.~S. and {De Vicente}, J. and {Desai}, S. and {Diehl}, H.~T. and {Doel}, P. and {Everett}, S. and {Ferrero}, I. and {Friedel}, D. and {Frieman}, J. and {Garc{\'\i}a-Bellido}, J. and {Gerdes}, D.~W. and {Gruen}, D. and {Gruendl}, R.~A. and {Gutierrez}, G. and {Hinton}, S.~R. and {Hollowood}, D.~L. and {Honscheid}, K. and {James}, D.~J. and {Kent}, S. and {Kuehn}, K. and {Kuropatkin}, N. and {Mena-Fern{\'a}ndez}, J. and {Miquel}, R. and {Ogando}, R.~L.~C. and {Palmese}, A. and {Pieres}, A. and {Malag{\'o}n}, A.~A. Plazas and {Raveri}, M. and {Reil}, K. and {Rodriguez-Monroy}, M. and {Sanchez}, E. and {Scarpine}, V. and {Sevilla-Noarbe}, I. and {Smith}, M. and {Suchyta}, E. and {Tarle}, G. and {To}, C. and {Weaverdyck}, N. and {DES Collaboration}},
        title = "{The Dark Energy Survey Supernova Program: Corrections on Photometry Due to Wavelength-dependent Atmospheric Effects}",
      journal = {\aj},
         year = 2023,
        month = jun,
       volume = {165},
       number = {6},
          eid = {222},
        pages = {222},
          doi = {10.3847/1538-3881/acca15},
archivePrefix = {arXiv},
       eprint = {2304.01858},
 primaryClass = {astro-ph.CO},
       adsurl = {https://ui.adsabs.harvard.edu/abs/2023AJ....165..222L}
}

@ARTICLE{Kessler2009,
       author = {{Kessler}, Richard and {Bernstein}, Joseph P. and {Cinabro}, David and {Dilday}, Benjamin and {Frieman}, Joshua A. and {Jha}, Saurabh and {Kuhlmann}, Stephen and {Miknaitis}, Gajus and {Sako}, Masao and {Taylor}, Matt and {Vanderplas}, Jake},
        title = "{SNANA: A Public Software Package for Supernova Analysis}",
      journal = {\pasp},
         year = 2009,
        month = sep,
       volume = {121},
       number = {883},
        pages = {1028},
          doi = {10.1086/605984},
archivePrefix = {arXiv},
       eprint = {0908.4280},
 primaryClass = {astro-ph.CO},
       adsurl = {https://ui.adsabs.harvard.edu/abs/2009PASP..121.1028K}
}

@ARTICLE{Hinton2020,
       author = {{Hinton}, Samuel and {Brout}, Dillon},
        title = "{Pippin: A pipeline for supernova cosmology}",
      journal = {The Journal of Open Source Software},
         year = 2020,
        month = mar,
       volume = {5},
       number = {47},
          eid = {2122},
        pages = {2122},
          doi = {10.21105/joss.02122},
       adsurl = {https://ui.adsabs.harvard.edu/abs/2020JOSS....5.2122H}
}

@ARTICLE{Riess1998,
       author = {{Riess}, Adam G. and {Filippenko}, Alexei V. and {Challis}, Peter and {Clocchiatti}, Alejandro and {Diercks}, Alan and {Garnavich}, Peter M. and {Gilliland}, Ron L. and {Hogan}, Craig J. and {Jha}, Saurabh and {Kirshner}, Robert P. and {Leibundgut}, B. and {Phillips}, M.~M. and {Reiss}, David and {Schmidt}, Brian P. and {Schommer}, Robert A. and {Smith}, R. Chris and {Spyromilio}, J. and {Stubbs}, Christopher and {Suntzeff}, Nicholas B. and {Tonry}, John},
        title = "{Observational Evidence from Supernovae for an Accelerating Universe and a Cosmological Constant}",
      journal = {\aj},
         year = 1998,
        month = sep,
       volume = {116},
       number = {3},
        pages = {1009-1038},
          doi = {10.1086/300499},
archivePrefix = {arXiv},
       eprint = {astro-ph/9805201},
 primaryClass = {astro-ph},
       adsurl = {https://ui.adsabs.harvard.edu/abs/1998AJ....116.1009R}
}

@ARTICLE{Perlmutter1999,
       author = {{Perlmutter}, S. and {Aldering}, G. and {Goldhaber}, G. and {Knop}, R.~A. and {Nugent}, P. and {Castro}, P.~G. and {Deustua}, S. and {Fabbro}, S. and {Goobar}, A. and {Groom}, D.~E. and {Hook}, I.~M. and {Kim}, A.~G. and {Kim}, M.~Y. and {Lee}, J.~C. and {Nunes}, N.~J. and {Pain}, R. and {Pennypacker}, C.~R. and {Quimby}, R. and {Lidman}, C. and {Ellis}, R.~S. and {Irwin}, M. and {McMahon}, R.~G. and {Ruiz-Lapuente}, P. and {Walton}, N. and {Schaefer}, B. and {Boyle}, B.~J. and {Filippenko}, A.~V. and {Matheson}, T. and {Fruchter}, A.~S. and {Panagia}, N. and {Newberg}, H.~J.~M. and {Couch}, W.~J. and {Project}, The Supernova Cosmology},
        title = "{Measurements of {\ensuremath{\Omega}} and {\ensuremath{\Lambda}} from 42 High-Redshift Supernovae}",
      journal = {\apj},
         year = 1999,
        month = jun,
       volume = {517},
       number = {2},
        pages = {565-586},
          doi = {10.1086/307221},
archivePrefix = {arXiv},
       eprint = {astro-ph/9812133},
 primaryClass = {astro-ph},
       adsurl = {https://ui.adsabs.harvard.edu/abs/1999ApJ...517..565P}
}

@ARTICLE{Switzer2025,
       author = {{Switzer}, Eric R. and {Bray}, Evan and {Will}, Scott D. and {Cromey}, Benjamin and {Gao}, Guangjun and {Groff}, Tyler D. and {Jurling}, Alden S. and {Kruk}, Jeffrey and {Marx}, Catherine T. and {Morey}, Peter A. and {Patel}, Jessica and {Quijada}, Manuel A. and {Rizzo}, Maxime J. and {Schlieder}, Joshua E. and {Wollack}, Edward J.},
        title = "{Laboratory characterization of widefield filter transmission for the Nancy Grace Roman Space Telescope's Wide Field Instrument}",
      journal = {\ao},
         year = 2025,
        month = dec,
       volume = {64},
       number = {35},
        pages = {10525},
          doi = {10.1364/AO.569503},
       adsurl = {https://ui.adsabs.harvard.edu/abs/2025ApOpt..6410525S}
}

@ARTICLE{Kessler2017,
       author = {{Kessler}, R. and {Scolnic}, D.},
        title = "{Correcting Type Ia Supernova Distances for Selection Biases and Contamination in Photometrically Identified Samples}",
      journal = {\apj},
         year = 2017,
        month = feb,
       volume = {836},
       number = {1},
          eid = {56},
        pages = {56},
          doi = {10.3847/1538-4357/836/1/56},
archivePrefix = {arXiv},
       eprint = {1610.04677},
 primaryClass = {astro-ph.CO},
       adsurl = {https://ui.adsabs.harvard.edu/abs/2017ApJ...836...56K}
}

@ARTICLE{Taylor2023,
       author = {{Taylor}, G. and {Jones}, D.~O. and {Popovic}, B. and {Vincenzi}, M. and {Kessler}, R. and {Scolnic}, D. and {Dai}, M. and {Kenworthy}, W.~D. and {Pierel}, J.~D.~R.},
        title = "{SALT2 versus SALT3: updated model surfaces and their impacts on type Ia supernova cosmology}",
      journal = {\mnras},
     year = 2023,
     month = apr,
     volume = {520},
     number = {4},
     pages = {5209-5224},
     doi = {10.1093/mnras/stad320},
archivePrefix = {arXiv},
       eprint = {2301.10644},
 primaryClass = {astro-ph.CO},
       adsurl = {https://ui.adsabs.harvard.edu/abs/2023MNRAS.520.5209T}
}

@UNPUBLISHED{Paulin2026,
       author = {Jillian Paulin and Masao Sako and et al.} ,
       title = "{Sundial: Simulating the Roman High-Latitude Time Domain Survey with the Roman Time Allocation Committee Observing Strategy}",
       note = {},
       year = {in prep},
}

@ARTICLE{Popovic2021,
       author = {{Popovic}, Brodie and {Brout}, Dillon and {Kessler}, Richard and {Scolnic}, Dan and {Lu}, Lisa},
        title = "{Improved Treatment of Host-galaxy Correlations in Cosmological Analyses with Type Ia Supernovae}",
      journal = {\apj},
         year = 2021,
        month = may,
       volume = {913},
       number = {1},
          eid = {49},
        pages = {49},
          doi = {10.3847/1538-4357/abf14f},
archivePrefix = {arXiv},
       eprint = {2102.01776},
 primaryClass = {astro-ph.CO},
       adsurl = {https://ui.adsabs.harvard.edu/abs/2021ApJ...913...49P}
}

@ARTICLE{Kenworthy2021,
       author = {{Kenworthy}, W.~D. and {Jones}, D.~O. and {Dai}, M. and {Kessler}, R. and {Scolnic}, D. and {Brout}, D. and {Siebert}, M.~R. and {Pierel}, J.~D.~R. and {Dettman}, K.~G. and {Dimitriadis}, G. and {Foley}, R.~J. and {Jha}, S.~W. and {Pan}, Y.-C. and {Riess}, A. and {Rodney}, S. and {Rojas-Bravo}, C.},
       title = "{SALT3: An Improved Type Ia Supernova Model for Measuring Cosmic Distances}",
       journal = {\apj},
       year = 2021,
       month = dec,
       volume = {923},
       number = {2},
       eid = {265},
       pages = {265},
       doi = {10.3847/1538-4357/ac30d8},
       archivePrefix = {arXiv},
       eprint = {2104.07795},
       primaryClass = {astro-ph.CO},
       adsurl = {https://ui.adsabs.harvard.edu/abs/2021ApJ...923..265K}
}

@ARTICLE{Pickles1998,
   author = {{Pickles}, A.~J.},
    title = "{A Stellar Spectral Flux Library: 1150-25000 {\AA}}",
  journal = {\pasp},
     year = 1998,
    month = jul,
   volume = {110},
    pages = {863-878},
      doi = {10.1086/316197}
}

@ARTICLE{Fitzpatrick1999,
       author = {{Fitzpatrick}, Edward L.},
        title = "{Correcting for the Effects of Interstellar Extinction}",
      journal = {\pasp},
         year = 1999,
        month = jan,
       volume = {111},
       number = {755},
        pages = {63-75},
          doi = {10.1086/316293},
archivePrefix = {arXiv},
       eprint = {astro-ph/9809387},
 primaryClass = {astro-ph},
       adsurl = {https://ui.adsabs.harvard.edu/abs/1999PASP..111...63F}
}

@ARTICLE{Bohlin2014,
       author = {{Bohlin}, R.~C. and {Gordon}, K.~D. and {Tremblay}, P.-E.},
        title = "{Techniques and Review of Absolute Flux Calibration from the Ultraviolet to the Mid-Infrared}",
      journal = {\pasp},
         year = 2014,
        month = jul,
       volume = {126},
       number = {941},
        pages = {711},
}

@ARTICLE{Marriner2011,
       author = {{Marriner}, John and {Bernstein}, J.~P. and {Kessler}, Richard and {Lampeitl}, Hubert and {Miquel}, Ramon and {Mosher}, Jennifer and {Nichol}, Robert C. and {Sako}, Masao and {Schneider}, Donald P. and {Smith}, Mathew},
        title = "{A More General Model for the Intrinsic Scatter in Type Ia Supernova Distance Moduli}",
      journal = {\apj},
         year = 2011,
        month = oct,
       volume = {740},
       number = {2},
          eid = {72},
        pages = {72},
          doi = {10.1088/0004-637X/740/2/72},
archivePrefix = {arXiv},
       eprint = {1107.4631},
 primaryClass = {astro-ph.CO},
       adsurl = {https://ui.adsabs.harvard.edu/abs/2011ApJ...740...72M}
}

@ARTICLE{Scolnic2015,
       author = {{Scolnic}, D. and {Casertano}, S. and {Riess}, A. and {Rest}, A. and {Schlafly}, E. and {Foley}, R.~J. and {Finkbeiner}, D. and {Tang}, C. and {Burgett}, W.~S. and {Chambers}, K.~C. and {Draper}, P.~W. and {Flewelling}, H. and {Hodapp}, K.~W. and {Huber}, M.~E. and {Kaiser}, N. and {Kudritzki}, R.~P. and {Magnier}, E.~A. and {Metcalfe}, N. and {Stubbs}, C.~W.},
        title = "{Supercal: Cross-calibration of Multiple Photometric Systems to Improve Cosmological Measurements with Type Ia Supernovae}",
      journal = {\apj},
         year = 2015,
        month = dec,
       volume = {815},
       number = {2},
          eid = {117},
        pages = {117},
          doi = {10.1088/0004-637X/815/2/117},
archivePrefix = {arXiv},
       eprint = {1508.05361},
 primaryClass = {astro-ph.IM},
       adsurl = {https://ui.adsabs.harvard.edu/abs/2015ApJ...815..117S}
}

@ARTICLE{Schlafly2012,
       author = {{Schlafly}, E.~F. and {Finkbeiner}, D.~P. and {Juri{\'c}}, M. and {Magnier}, E.~A. and {Burgett}, W.~S. and {Chambers}, K.~C. and {Grav}, T. and {Hodapp}, K.~W. and {Kaiser}, N. and {Kudritzki}, R.-P. and {Martin}, N.~F. and {Morgan}, J.~S. and {Price}, P.~A. and {Rix}, H.-W. and {Stubbs}, C.~W. and {Tonry}, J.~L. and {Wainscoat}, R.~J.},
        title = "{Photometric Calibration of the First 1.5 Years of the Pan-STARRS1 Survey}",
      journal = {\apj},
         year = 2012,
        month = sep,
       volume = {756},
       number = {2},
          eid = {158},
        pages = {158},
          doi = {10.1088/0004-637X/756/2/158},
archivePrefix = {arXiv},
       eprint = {1201.2208},
 primaryClass = {astro-ph.IM},
       adsurl = {https://ui.adsabs.harvard.edu/abs/2012ApJ...756..158S}
}

@ARTICLE{Astropy2018,
       author = {{Astropy Collaboration} and {Price-Whelan}, A.~M. and {Sip{\H{o}}cz}, B.~M. and {G{\"u}nther}, H.~M. and {Lim}, P.~L. and {Crawford}, S.~M. and {Conseil}, S. and {Shupe}, D.~L. and {Craig}, M.~W. and {Dencheva}, N. and {Ginsburg}, A. and {VanderPlas}, J.~T. and {Bradley}, L.~D. and {P{\'e}rez-Su{\'a}rez}, D. and {de Val-Borro}, M. and {Aldcroft}, T.~L. and {Cruz}, K.~L. and {Robitaille}, T.~P. and {Tollerud}, E.~J. and {Ardelean}, C. and {Babej}, T. and {Bach}, Y.~P. and {Bachetti}, M. and {Bakanov}, A.~V. and {Bamford}, S.~P. and {Barentsen}, G. and {Barmby}, P. and {Baumbach}, A. and {Berry}, K.~L. and {Biscani}, F. and {Boquien}, M. and {Bostroem}, K.~A. and {Bouma}, L.~G. and {Brammer}, G.~B. and {Bray}, E.~M. and {Breytenbach}, H. and {Buddelmeijer}, H. and {Burke}, D.~J. and {Calderone}, G. and {Cano Rodr{\'\i}guez}, J.~L. and {Cara}, M. and {Cardoso}, J.~V.~M. and {Cheedella}, S. and {Copin}, Y. and {Corrales}, L. and {Crichton}, D. and {D'Avella}, D. and {Deil}, C. and {Depagne}, {\'E}. and {Dietrich}, J.~P. and {Donath}, A. and {Droettboom}, M. and {Earl}, N. and {Erben}, T. and {Fabbro}, S. and {Ferreira}, L.~A. and {Finethy}, T. and {Fox}, R.~T. and {Garrison}, L.~H. and {Gibbons}, S.~L.~J. and {Goldstein}, D.~A. and {Gommers}, R. and {Greco}, J.~P. and {Greenfield}, P. and {Groener}, A.~M. and {Grollier}, F. and {Hagen}, A. and {Hirst}, P. and {Homeier}, D. and {Horton}, A.~J. and {Hosseinzadeh}, G. and {Hu}, L. and {Hunkeler}, J.~S. and {Ivezi{\'c}}, {\v{Z}}. and {Jain}, A. and {Jenness}, T. and {Kanarek}, G. and {Kendrew}, S. and {Kern}, N.~S. and {Kerzendorf}, W.~E. and {Khvalko}, A. and {King}, J. and {Kirkby}, D. and {Kulkarni}, A.~M. and {Kumar}, A. and {Lee}, A. and {Lenz}, D. and {Littlefair}, S.~P. and {Ma}, Z. and {Macleod}, D.~M. and {Mastropietro}, M. and {McCully}, C. and {Montagnac}, S. and {Morris}, B.~M. and {Mueller}, M. and {Mumford}, S.~J. and {Muna}, D. and {Murphy}, N.~A. and {Nelson}, S. and {Nguyen}, G.~H. and {Ninan}, J.~P. and {N{\"o}the}, M. and {Ogaz}, S. and {Oh}, S. and {Parejko}, J.~K. and {Parley}, N. and {Pascual}, S. and {Patil}, R. and {Patil}, A.~A. and {Plunkett}, A.~L. and {Prochaska}, J.~X. and {Rastogi}, T. and {Reddy Janga}, V. and {Sabater}, J. and {Sakurikar}, P. and {Seifert}, M. and {Sherbert}, L.~E. and {Sherwood-Taylor}, H. and {Shih}, A.~Y. and {Sick}, J. and {Silbiger}, M.~T. and {Singanamalla}, S. and {Singer}, L.~P. and {Sladen}, P.~H. and {Sooley}, K.~A. and {Sornarajah}, S. and {Streicher}, O. and {Teuben}, P. and {Thomas}, S.~W. and {Tremblay}, G.~R. and {Turner}, J.~E.~H. and {Terr{\'o}n}, V. and {van Kerkwijk}, M.~H. and {de la Vega}, A. and {Watkins}, L.~L. and {Weaver}, B.~A. and {Whitmore}, J.~B. and {Woillez}, J. and {Zabalza}, V. and {Astropy Contributors}},
        title = "{The Astropy Project: Building an Open-science Project and Status of the v2.0 Core Package}",
      journal = {\aj},
         year = 2018,
        month = sep,
       volume = {156},
       number = {3},
          eid = {123},
        pages = {123},
          doi = {10.3847/1538-3881/aabc4f},
archivePrefix = {arXiv},
       eprint = {1801.02634},
 primaryClass = {astro-ph.IM},
       adsurl = {https://ui.adsabs.harvard.edu/abs/2018AJ....156..123A}
}

@ARTICLE{2020SciPy-NMeth,
      author  = {Virtanen, Pauli and Gommers, Ralf and Oliphant, Travis E. and
                Haberland, Matt and Reddy, Tyler and Cournapeau, David and
                Burovski, Evgeni and Peterson, Pearu and Weckesser, Warren and
                Bright, Jonathan and {van der Walt}, St{\'e}fan J. and
                Brett, Matthew and Wilson, Joshua and Millman, K. Jarrod and
                Mayorov, Nikolay and Nelson, Andrew R. J. and Jones, Eric and
                Kern, Robert and Larson, Eric and Carey, C J and
                Polat, {\.I}lhan and Feng, Yu and Moore, Eric W. and
                {VanderPlas}, Jake and Laxalde, Denis and Perktold, Josef and
                Cimrman, Robert and Henriksen, Ian and Quintero, E. A. and
                Harris, Charles R. and Archibald, Anne M. and
                Ribeiro, Ant{\^o}nio H. and Pedregosa, Fabian and
                {van Mulbregt}, Paul and {SciPy 1.0 Contributors}},
      title   = {{{SciPy} 1.0: Fundamental Algorithms for Scientific
                Computing in Python}},
      journal = {Nature Methods},
      year    = {2020},
      volume  = {17},
      pages   = {261--272},
      adsurl  = {https://rdcu.be/b08Wh},
      doi     = {10.1038/s41592-019-0686-2},
}

@ARTICLE{Hinton2016,
      author = {{Hinton}, S.~R.},
       title = "{ChainConsumer}",
     journal = {The Journal of Open Source Software},
        year = 2016,
       month = aug,
      volume = 1,
         eid = {00045},
       pages = {00045},
         doi = {10.21105/joss.00045},
      adsurl = {http://adsabs.harvard.edu/abs/2016JOSS....1...45H},
   }

@Article{Hunter:2007,
  Author    = {Hunter, J. D.},
  Title     = {Matplotlib: A 2D graphics environment},
  Journal   = {Computing in Science \& Engineering},
  Volume    = {9},
  Number    = {3},
  Pages     = {90--95},
  publisher = {IEEE COMPUTER SOC},
  doi       = {10.1109/MCSE.2007.55},
  year      = 2007
}

@ARTICLE{Popovic2025,
       author = {{Popovic}, B. and {Rigault}, M. and {Smith}, M. and {Ginolin}, M. and {Goobar}, A. and {Kenworthy}, W.~D. and {Ganot}, C. and {Ruppin}, F. and {Dimitriadis}, G. and {Johansson}, J. and {Amenouche}, M. and {Aubert}, M. and {Barjou-Delayre}, C. and {Burgaz}, U. and {Carreres}, B. and {Feinstein}, F. and {Fouchez}, D. and {Galbany}, L. and {de Jaeger}, T. and {Kim}, Y.-L. and {Lacroix}, L. and {Nugent}, P.~E. and {Racine}, B. and {Rosselli}, D. and {Rosnet}, P. and {Sollerman}, J. and {Hale}, D. and {Laher}, R. and {M{\"u}ller-Bravo}, T.~E. and {Reed}, R. and {Rusholme}, B. and {Terwel}, J.},
        title = "{ZTF SN Ia DR2: Evidence of changing dust distribution with redshift using type Ia supernovae}",
      journal = {\aap},
         year = 2025,
        month = feb,
       volume = {694},
          eid = {A5},
        pages = {A5},
          doi = {10.1051/0004-6361/202450391},
archivePrefix = {arXiv},
       eprint = {2406.06215},
 primaryClass = {astro-ph.CO},
       adsurl = {https://ui.adsabs.harvard.edu/abs/2025A&A...694A...5P}
}

@ARTICLE{Oke1974,
       author = {{Oke}, J.~B.},
        title = "{Absolute Spectral Energy Distributions for White Dwarfs}",
      journal = {\apjs},
         year = 1974,
        month = feb,
       volume = {27},
        pages = {21},
          doi = {10.1086/190287},
       adsurl = {https://ui.adsabs.harvard.edu/abs/1974ApJS...27...21O}
}

@techreport{Williams2025RomanCalibration,
            author={Williams, Benjamin and Bellini, Andrea and Walth, Gregory and Casertano, Stefano and Zimmerman, Neil T. and Calamida, Annalisa and Gennaro, Mario and Kruk, Jeffrey W. and Mehta, Vihang and {Roman Wide Field Instrument Calibration Working Group}},
            title= {{Nancy Grace Roman Space Telescope: Wide Field Instrument Calibration Touchstone Field Recommendations}},
  year         = {2025},
  month        = mar,
  day          = {31},
  url = {https://roman.gsfc.nasa.gov/science/calibration/WFI_Touchstone_Fields_RevA_2025-03-31.pdf}
}

@ARTICLE{Mosby2020,
       author = {{Mosby}, Gregory and {Rauscher}, Bernard J. and {Bennett}, Chris and {Cheng}, Edward S. and {Cheung}, Stephanie and {Cillis}, Analia and {Content}, David and {Cottingham}, Dave and {Foltz}, Roger and {Gygax}, John and {Hill}, Robert J. and {Kruk}, Jeffrey W. and {Mah}, Jon and {Meier}, Lane and {Merchant}, Chris and {Miko}, Laddawan and {Piquette}, Eric C. and {Waczynski}, Augustyn and {Wen}, Yiting},
       title = "{Properties and characteristics of the Nancy Grace Roman Space Telescope H4RG-10 detectors}", journal = {Journal of Astronomical Telescopes, Instruments, and Systems},
       year = 2020,
       month = oct,
       volume = {6},
          eid = {046001},
        pages = {046001},
          doi = {10.1117/1.JATIS.6.4.046001},
          archivePrefix = {arXiv},
       eprint = {2005.00505},
 primaryClass = {astro-ph.IM},
       adsurl = {https://ui.adsabs.harvard.edu/abs/2020JATIS...6d6001M}
}

@ARTICLE{Aldoroty2026,
       author = {{Aldoroty}, L. and {Scolnic}, D. and {Kannawadi}, A. and {Knop}, R.~A. and {Rose}, B.~M. and {Hounsell}, R. and {Troxel}, M.~A. and {The Roman Supernova Project Infrastructure Team}},
        title = "{Initial Characterization of Stellar Photometry of Roman Images from the OpenUniverse Simulations}",
      journal = {\aj},
         year = 2026,
        month = mar,
       volume = {171},
       number = {3},
          eid = {129},
        pages = {129},
          doi = {10.3847/1538-3881/ae3714},
archivePrefix = {arXiv},
       eprint = {2506.04332},
 primaryClass = {astro-ph.IM},
       adsurl = {https://ui.adsabs.harvard.edu/abs/2026AJ....171..129A}
}

@ARTICLE{Pierel2022,
       author = {{Pierel}, J.~D.~R. and {Jones}, D.~O. and {Kenworthy}, W.~D. and {Dai}, M. and {Kessler}, R. and {Ashall}, C. and {Do}, A. and {Peterson}, E.~R. and {Shappee}, B.~J. and {Siebert}, M.~R. and {Barna}, T. and {Brink}, T.~G. and {Burke}, J. and {Calamida}, A. and {Camacho-Neves}, Y. and {de Jaeger}, T. and {Filippenko}, A.~V. and {Foley}, R.~J. and {Galbany}, L. and {Fox}, O.~D. and {Gomez}, S. and {Hiramatsu}, D. and {Hounsell}, R. and {Howell}, D.~A. and {Jha}, S.~W. and {Kwok}, L.~A. and {P{\'e}rez-Fournon}, I. and {Poidevin}, F. and {Rest}, A. and {Rubin}, D. and {Scolnic}, D.~M. and {Shirley}, R. and {Strolger}, L.~G. and {Tinyanont}, S. and {Wang}, Q.},
        title = "{SALT3-NIR: Taking the Open-source Type Ia Supernova Model to Longer Wavelengths for Next-generation Cosmological Measurements}",
      journal = {\apj},
         year = 2022,
        month = nov,
       volume = {939},
       number = {1},
          eid = {11},
        pages = {11},
          doi = {10.3847/1538-4357/ac93f9},
archivePrefix = {arXiv},
       eprint = {2209.05594},
 primaryClass = {astro-ph.CO},
       adsurl = {https://ui.adsabs.harvard.edu/abs/2022ApJ...939...11P}
}

@ARTICLE{Marlin2025,
       author = {{Marlin}, Elijah G. and {Murakami}, Yukei S. and {Brout}, Dillon and {Tweddle}, Jack W. and {Popovic}, Brodie and {Smith}, Ken W. and {Smartt}, Stephen J. and {Scolnic}, Daniel M. and {Jones}, David and {Peterson}, Erik R. and {Riess}, Adam G. and {Vincenzi}, Maria and {Sherman}, Nora F. and {Acevedo}, Maria and {Milstein}, Jasper and {Dixon}, Mitchell and {Rest}, Armin},
        title = "{TITAN DR1: An Improved, Validated, and Systematically-Controlled Recalibration of ATLAS Photometry toward Type Ia Supernova Cosmology}",
      journal = {arXiv e-prints},
         year = 2025,
        month = dec,
          eid = {arXiv:2512.21903},
        pages = {arXiv:2512.21903},
          doi = {10.48550/arXiv.2512.21903},
archivePrefix = {arXiv},
       eprint = {2512.21903},
 primaryClass = {astro-ph.CO},
       adsurl = {https://ui.adsabs.harvard.edu/abs/2025arXiv251221903M}
}

@ARTICLE{Troxel2025,
       author = {{OpenUniverse} and {LSST Dark Energy Science Collaboration} and {Roman HLIS Project Infrastructure} and {Roman Rapid Project Infrastructure Team} and {Roman Supernova Cosmology Project Infrastructure Team} and {Alarcon}, A. and {Aldoroty}, L. and {Beltz-Mohrmann}, G. and {Bera}, A. and {Blazek}, J. and {Bogart}, J. and {Braeunlich}, G. and {Broughton}, A. and {Cao}, K. and {Chiang}, J. and {Chisari}, N.~E. and {Desai}, V. and {Fang}, Y. and {Galbany}, L. and {Hearin}, A. and {Heitmann}, K. and {Hirata}, C. and {Hounsell}, R. and {Jain}, B. and {Jarvis}, M. and {Jencson}, J. and {Kannawadi}, A. and {Kasliwal}, M.~K. and {Kessler}, R. and {Kiessling}, A. and {Knop}, R. and {Kovacs}, E. and {Laher}, R. and {Laliotis}, K. and {Lin}, C. and {Lopes}, I. and {MacBeth}, E. and {Mahabal}, A. and {Mandelbaum}, R. and {Masiero}, J. and {Mau}, S. and {Meehan}, C. and {Meyers}, J. and {Moraes}, B. and {Paladini}, R. and {Pearl}, A. and {Malagon}, A. Plazas and {Rose}, B. and {Rubin}, D. and {Rusholme}, B. and {Santos}, A. and {{\v{S}}ar{\v{c}}evi{\'c}}, N. and {Scolnic}, D. and {Singhal}, J. and {Troxel}, M.~A. and {van Alfen}, N. and {van Dyke}, S. and {Walter}, C.~W. and {Wu}, T. and {Yamamoto}, M. and {Yan}, L. and {Zhang}, T.},
        title = "{OpenUniverse2024: a shared, simulated view of the sky for the next generation of cosmological surveys}",
      journal = {\mnras},
         year = 2025,
        month = dec,
       volume = {544},
       number = {4},
        pages = {3799-3823},
          doi = {10.1093/mnras/staf1833},
archivePrefix = {arXiv},
       eprint = {2501.05632},
 primaryClass = {astro-ph.CO},
       adsurl = {https://ui.adsabs.harvard.edu/abs/2025MNRAS.544.3799O}
}

@ARTICLE{Brout2021,
       author = {{Brout}, Dillon and {Hinton}, Samuel R. and {Scolnic}, Dan},
        title = "{Binning is Sinning (Supernova Version): The Impact of Self-calibration in Cosmological Analyses with Type Ia Supernovae}",
      journal = {\apjl},
         year = 2021,
        month = may,
       volume = {912},
       number = {2},
          eid = {L26},
        pages = {L26},
          doi = {10.3847/2041-8213/abf4db},
archivePrefix = {arXiv},
       eprint = {2012.05900},
 primaryClass = {astro-ph.CO},
       adsurl = {https://ui.adsabs.harvard.edu/abs/2021ApJ...912L..26B}
}

@ARTICLE{Kunz2007,
       author = {{Kunz}, Martin and {Bassett}, Bruce A. and {Hlozek}, Ren{\'e}e A.},
        title = "{Bayesian estimation applied to multiple species}",
      journal = {\prd},
         year = 2007,
        month = may,
       volume = {75},
       number = {10},
          eid = {103508},
        pages = {103508},
          doi = {10.1103/PhysRevD.75.103508},
archivePrefix = {arXiv},
       eprint = {astro-ph/0611004},
 primaryClass = {astro-ph},
       adsurl = {https://ui.adsabs.harvard.edu/abs/2007PhRvD..75j3508K}
}

@ARTICLE{Hlozek2012,
       author = {{Hlozek}, Ren{\'e}e and {Kunz}, Martin and {Bassett}, Bruce and {Smith}, Mat and {Newling}, James and {Varughese}, Melvin and {Kessler}, Rick and {Bernstein}, Joseph P. and {Campbell}, Heather and {Dilday}, Ben and {Falck}, Bridget and {Frieman}, Joshua and {Kuhlmann}, Steve and {Lampeitl}, Hubert and {Marriner}, John and {Nichol}, Robert C. and {Riess}, Adam G. and {Sako}, Masao and {Schneider}, Donald P.},
        title = "{Photometric Supernova Cosmology with BEAMS and SDSS-II}",
      journal = {\apj},
         year = 2012,
        month = jun,
       volume = {752},
       number = {2},
          eid = {79},
        pages = {79},
          doi = {10.1088/0004-637X/752/2/79},
archivePrefix = {arXiv},
       eprint = {1111.5328},
 primaryClass = {astro-ph.CO},
       adsurl = {https://ui.adsabs.harvard.edu/abs/2012ApJ...752...79H}
}

@article{Wang2008,
  author  = {Wang, Y.},
  title   = {Figure of Merit for Dark Energy Constraints 
             from Current Observational Data},
  journal = {Phys. Rev. D},
  volume  = {77},
  pages   = {123525},
  year    = {2008}
}

@ARTICLE{SFD1998,
       author = {{Schlegel}, David J. and {Finkbeiner}, Douglas P. and {Davis}, Marc},
        title = "{Maps of Dust Infrared Emission for Use in Estimation of Reddening and Cosmic Microwave Background Radiation Foregrounds}",
      journal = {\apj},
         year = 1998,
        month = jun,
       volume = {500},
       number = {2},
        pages = {525-553},
          doi = {10.1086/305772},
archivePrefix = {arXiv},
       eprint = {astro-ph/9710327},
 primaryClass = {astro-ph},
       adsurl = {https://ui.adsabs.harvard.edu/abs/1998ApJ...500..525S}
}

@ARTICLE{Popovic2023,
       author = {{Popovic}, Brodie and {Brout}, Dillon and {Kessler}, Richard and {Scolnic}, Daniel},
        title = "{The Pantheon+ Analysis: Forward Modeling the Dust and Intrinsic Color Distributions of Type Ia Supernovae, and Quantifying Their Impact on Cosmological Inferences}",
      journal = {\apj},
         year = 2023,
        month = mar,
       volume = {945},
       number = {1},
          eid = {84},
        pages = {84},
          doi = {10.3847/1538-4357/aca273},
archivePrefix = {arXiv},
       eprint = {2112.04456},
 primaryClass = {astro-ph.CO},
       adsurl = {https://ui.adsabs.harvard.edu/abs/2023ApJ...945...84P}
}

\end{document}